# Very Local Interstellar Medium Revealed by Complete Solar Cycle of Interstellar Neutral Helium Observations with IBEX

P. Swaczyna[1,*], M. A. Kubiak[2], M. Bzowski[2], J. Bower[3], S. A. Fuselier[4,5], A. Galli[6],
D. Heirtzler[3], D. J. McComas[1], E. Möbius[3], F. Rahmanifard[3], N. A. Schwadron[1,3]

[1]Department of Astrophysical Sciences, Princeton University, Princeton, NJ 08544, USA
[2]Space Research Centre PAS (CBK PAN), Bartycka 18a, 00-716 Warsaw, Poland
[3]Physics Department, Space Science Center, University of New Hampshire, Durham, NH 03824, USA
[4]Southwest Research Institute, San Antonio, TX 78228, USA
[5]University of Texas at San Antonio, San Antonio, TX, USA
[6]Physics Institute, University of Bern, Bern, 3012, Switzerland

## Abstract

The IBEX-Lo instrument on board the Interstellar Boundary Explorer (IBEX) mission samples interstellar neutral (ISN) helium atoms penetrating the heliosphere from the very local interstellar medium (VLISM). In this study, we analyze the IBEX-Lo ISN helium observations covering a complete solar cycle, from 2009 through 2020 using a comprehensive uncertainty analysis including statistical and systematic sources. We employ the Warsaw Test Particle Model to simulate ISN helium fluxes at IBEX, which are subsequently compared with the observed count rate in the three lowest energy steps of IBEX-Lo. The $\chi^2$ analysis shows that the ISN helium flows from ecliptic $(\lambda, \beta) = (255.59° \pm 0.23°, 5.14° \pm 0.08°)$, with speed $v_{\text{HP}} = 25.86 \pm 0.21$ km s$^{-1}$ and temperature $T_{\text{HP}} = 7450 \pm 140$ K at the heliopause. Accounting for gravitational attraction and elastic collisions, the ISN helium speed and temperature in the pristine VLISM far from the heliopause are $v_{\text{VLISM}} = 25.9$ km s$^{-1}$ and $T_{\text{VLISM}} = 6150$ K, respectively. The time evolution of the ISN helium fluxes at 1 au over 12 years suggests significant changes in the IBEX-Lo detection efficiency, higher ionization rates of ISN helium atoms in the heliosphere than assumed in the model, or an additional unaccounted signal source in the analyzed observations. Nevertheless, we do not find any indication of the evolution of the derived parameters of ISN helium over the period analyzed. Finally, we argue that the continued operation of IBEX-Lo to overlap with the Interstellar Mapping and Acceleration Probe (IMAP) will be pivotal in tracking possible physical changes in the VLISM.

## 1. Introduction

The very local interstellar medium (VLISM) in the proximity of our Sun is filled with partially ionized, warm, and magnetized plasma (Frisch et al. 2011). The solar wind inflates the heliosphere in the VLISM, which keeps most of the interstellar ionized particles outside the heliopause (e.g., Parker 1961). Neutral atoms, however, flow through the heliospheric boundaries and are detected close to the Sun (Wallis 1975). Therefore, direct sampling of interstellar neutral (ISN) atoms provides a unique ground-truth measurement of the VLISM conditions in the proximity of the heliopause. The two most abundant neutral species are hydrogen and helium. While ISN hydrogen is significantly modulated outside the heliopause due to charge exchange collisions with interstellar protons (Izmodenov et al. 2001, 2004), most of ISN helium atoms originate from the pristine VLISM and thus are much less affected in the heliosphere. Moreover, ISN hydrogen atoms are strongly ionized inside the heliosphere (Bzowski et al. 2013; Sokół et al. 2019a), making them less abundant at 1 au than ISN helium (Sokół et al. 2019b). Consequently, ISN helium is the main species used to derive the VLISM flow and temperature beyond the heliopause.

---

[*] Corresponding author (swaczyna@princeton.edu)

The Interstellar Boundary Explorer (IBEX, McComas et al. 2009) mission is dedicated to making observations of neutral atoms. One of its two instruments, IBEX-Lo (Fuselier et al. 2009), observes ISN atoms in its lowest energy steps (Möbius et al. 2009a, 2009b). Analyses of IBEX-Lo observations from the first two years of operation (Bzowski et al. 2012; Möbius et al. 2012) suggested that the ISN He flow was slower than previously deduced from Ulysses/GAS observations (Witte 2004; Witte et al. 2004). These results raised the question of whether the VLISM conditions ahead of the heliosphere could change between the observation period of Ulysses and IBEX (Frisch et al. 2013, 2015; Lallement & Bertaux 2014). Reanalyzes of Ulysses data (Bzowski et al. 2014; Wood et al. 2015) and further analyses of IBEX data based on six years of observations (Bzowski et al. 2015; Leonard et al. 2015; Möbius et al. 2015a; Schwadron et al. 2015) showed that parameters from IBEX were consistent with those deduced from Ulysses (McComas et al. 2015a). Furthermore, Bzowski & Kubiak (2020) demonstrated that the travel time of ISN He atoms from the pristine VLISM to 1 au typically exceeds 30 years, with a significant time spread for individual atoms depending on their velocities relative to the Sun.

ISN helium atoms enter the IBEX-Lo field of view twice a year (Sokół et al. 2015a). In winter, IBEX-Lo, together with Earth, moves against the ISN helium flow, and the effective velocity in the instrument frame is a sum of these two speeds. In fall, however, the spacecraft moves with the ISN flow, and the effective energy is below the sputtering energy threshold (Galli et al. 2015), and atoms are not detected. Since IBEX-Lo does not measure the speeds of observed ISN helium atoms, faster atoms, whose trajectories are less affected by the solar gravity, are not distinguishable from slower atoms. This indistinguishability results in a tight correlation of the deduced flow vector components and temperature along a 4-dimensional correlation tube, but the uncertainty along the tube, especially from a limited dataset analyzed in the first studies, is much larger than across the tube (Lee et al. 2012, 2015; McComas et al. 2012, 2015a).

Even though most ISN helium atoms sampled at 1 au by IBEX are formed in the pristine VLISM, a small portion of these atoms charge exchange with interstellar $He^+$ ions, which result in the formation of the secondary population (Bzowski et al. 2017). This secondary population has the highest density ~200 au ahead of the Sun but at 1 au it is expected to be strongly attenuated by ballistic selection effects (Kubiak et al. 2019). The secondary population was discovered in the IBEX data and originally dubbed the Warm Breeze (Kubiak et al. 2014). Kubiak et al. (2016) showed that the apparent inflow direction of the Warm Breeze is aligned with the plane defined by the inflow direction and the interstellar magnetic field orientation. The deflection of the Warm Breeze ahead of the heliosphere has been confirmed by Fraternale et al. (2021) using a global heliosphere model that includes helium ions and atoms with appropriate charge exchange between these species. Moreover, they showed that the secondary population does not follow the Maxwell distribution at the heliopause. Finally, analysis of this population allowed for determining the $He^+$ ion density in the VLISM (Bzowski et al. 2019). The secondary helium population is a significant contribution to the IBEX signal and may influence the determination of the VLISM parameters (Möbius et al. 2015a; Swaczyna et al. 2015).

Wood et al. (2019) and Swaczyna et al. (2019a) pointed out that the IBEX observations suggest that the distribution function of the primary ISN helium in the pristine VLISM is not a Maxwellian. This finding indicates that the VLISM in the Sun's proximity is not in thermal equilibrium. In part, this may be because the primary ISN helium is modified by elastic collisions in the outer heliosheath (Chassefière & Bertaux 1987; Swaczyna et al. 2021) and in the solar wind (Gruntman 2018). Moreover, charge exchange collisions cause small momentum exchanges between colliding particles (Swaczyna et al. 2019b).

This paper analyzes the first 12 years of ISN helium observations from IBEX to find the VLISM parameters ahead of the heliosphere. Section 2 presents the IBEX-Lo data and uncertainty analysis used in this study. A brief description of the Warsaw Test Particle Model employed to simulate the ISN helium fluxes is presented in Section 3. The method of derivation of the best-fit VLISM parameters is provided in Section 4. The main results of this work are presented in Section 5. Section 6 discusses temporal changes to the ISN helium flux magnitude, and Section 7 shows stability of the obtained VLISM parameters over time. Finally, conclusions and summary are in Section 8.

## 2. IBEX-Lo Data

IBEX-Lo uses diamond-like carbon conversion surfaces to convert impacting neutral atoms to negative ions (Wieser et al. 2007). The energy of these converted ions is further selected using an electrostatic analyzer (ESA) (Fuselier et al. 2009). After exiting the ESA, ions are accelerated by a post-acceleration (PAC) voltage, and their species are identified using time-of-flight (TOF) measurements. This setup allows for unprecedented reduction of background, especially from the UV photons, which significantly influenced Ulysses/GAS observations. Helium atoms efficiently sputter $H^-$ ions from the conversion surface (Möbius et al. 2012) but direct charge-conversion produces only short-living metastable negative $He^-$ ions, and thus the probability of detection through direct conversion is low (Wurz et al. 2008). The sputtered ions form a wide energy spectrum, so complete reconstruction of the original atom energy is not possible.

In the winter peak, the helium atom speed in the IBEX frame is up to ~80 km s$^{-1}$, i.e., with an energy of ~134 eV (Swaczyna et al. 2018). This energy corresponds to ESA step 4. However, since the spectrum of ions produced by helium atoms is wide, they are observed in all four of the lowest energy steps. A precise calibration for helium atoms is not available, so the energy response must be estimated from the observations. In this study, we use ESA steps 1-3 because ESA step 4 shows significantly reduced ISN atom fluxes and a shift of the ISN longitudinal peak caused by more than a two-fold decrease of the response function in ESA step 4 for the energy range included here (Schwadron et al. 2022).

### 2.1 Orbit Selection

The IBEX-Lo boresight follows a great circle perpendicular to the spacecraft spin axis, which is periodically repointed to approximately follow the Sun. Initially, IBEX was placed on a highly elliptical orbit with an orbital period of ~7.5 days, and the spin axis was repointed once per orbit when the spacecraft was close to the perigee. In mid-2011, IBEX was moved to a long-term stable lunar resonance orbit with the orbital period of ~9.1 days (McComas et al. 2011). Since this change, the spin axis is repointed twice per orbit around perigee and apogee. Data are organized based on the spin axis pointing, and we refer to the IBEX orientations using orbit or arc identifiers.

The selection of the IBEX data for this study follows the criteria previously used by Bzowski et al. (2015) and Swaczyna et al. (2018). First, we select IBEX orbits and arcs when the ecliptic longitude of the IBEX spin axis is between 295° to 335°, which approximately corresponds to Earth's ecliptic longitude from 115° to 155°. This range follows the selection in Swaczyna et al. (2018) and is shorter by 5° (typically removing one orbit or arc in each year) from the one used by Bzowski et al. (2015) to avoid contamination from ISN hydrogen atoms (Galli et al. 2019). The selected observations were made in January and February of each year, which is the peak of the IBEX ISN "season". Swaczyna et al. (2015) found that reanalysis of the spacecraft attitude data is needed to obtain high precision spin axis pointing. We use these spin axis directions also for this present study and include them in the derived products (Appendix C). On average, about 5 orbits or 8 arcs are pointed within this ecliptic longitude range each year.

Early in the mission, a uniform cadence 1-2-3-4-5-6-7-8 of all eight energy steps was used for most of the orbits and arcs. Sometimes, one of two special modes is used, during which only two energy steps were sampled. The first one – the oxygen mode – uses special energy settings tuned to ISN oxygen for 7/8 of the total time, and ESA step 2 is observed during the remaining 1/8 of the time. This mode was used in 2011 and 2012 to enhance ISN oxygen observations. The second special mode is implemented for cross-calibration (X-cal) between the high and the low resolutions sections of the collimator (Fuselier et al. 2009). In this mode, ESA step 2 is also observed for 1/8 of the time, but for the remaining 7/8 of the time, IBEX-Lo is in the high-resolution mode. While we do not use the oxygen or high-resolution data in this study, data from ESA step 2 collected during these two special modes are included. If a special mode is applied for only a part of an orbit/arc, then we use the part with the normal stepping. In 2016, a new sweep table: 1-1-2-2-5-6-7-8 was implemented, meaning that the accumulation times in ESA 1 and 2 are doubled, but ESA 3 and 4 are skipped. Yet another sweep table was implemented during ISN seasons since 2017: 1-1-2-2-3-6-7-8, i.e., ESA 4 and 5 are skipped. The list of selected orbits for each period is provided in Table 1 together with the applied ESA sweep table. Note that ESA step 2 is always observed for the selected orbits/arcs, while availability ESA 1 or 3 may differ.

**Table 1**
**Orbit/arc Selected for This Study**

| Season | Orbits/arcs | ESA Sweep Table | Missing Orbits | ESA Sweep Table Exceptions |
|---|---|---|---|---|
| 2009 | 14, 15, 16, 17, 18 | 1-2-3-4-5-6-7-8 | | |
| 2010 | 63, 64, 65, 66 | 1-2-3-4-5-6-7-8 | 62: no data | |
| 2011 | 110, 112, 113, 114 | Oxygen mode | 111: spun | |
| 2012 | 153b, 154a,b, 156a,b, 157a | X-cal mode | 155a,b: no data | 156b, 157a: 1-2-3-4-5-6-7-8 |
| 2013 | 193a,b, 194a,b, 195a,b, 196a,b, 197a,b | 1-2-3-4-5-6-7-8 | | |
| 2014 | 233b, 234a,b, 235a,b, 236a,b, 237a | 1-2-3-4-5-6-7-8 | 237b: spun or high TOF 2 | |
| 2015 | 273b, 274b, 275a,b, 276a,b | 1-2-3-4-5-6-7-8 | 274a, 277a,b: high TOF 2 | 273b: X-cal |
| 2016 | 314a,b, 315a | 1-1-2-2-5-6-7-8 | ≥315b: star tracker anomaly | |
| 2017 | 354a,b, 355a,b, 356a,b, 357a,b | 1-1-2-2-3-6-7-8 | | 356a, 357a: X-cal, 354a: 1-2-3-4-5-6-7-8 |
| 2018 | 394b, 395a,b, 396a,b, 397a,b, 398a | 1-1-2-2-3-6-7-8 | | |
| 2019 | 434b, 435a,b, 436a,b, 437a,b, 438a | 1-1-2-2-3-6-7-8 | | 435a: 1-2-3-4-5-6-7-8 |
| 2020 | 474b, 475a,b, 476a,b, 477a,b, 478a | 1-1-2-2-3-6-7-8 | | 475a: 1-2-3-4-5-6-7-8 |

Table 1 also indicates reasons why some orbits and arcs are missing in the selected dataset even though they meet the pointing requirement. Specifically, orbits 62 and 155 (both arcs) are not available due to onboard computer reset causing loss of science data. Most of season 2016 is lost due to a star tracker anomaly that started in arc 315b. Since the proper operation of the star tracker was not restored until arc 325b, this season includes only 3 arcs. Five other orbits/arcs (111, 237b, 274a, 277a,b) do not include any times that meet the "good times" criteria specified in Section 2.2.

## 2.2 Data Selection

For our analysis, we use only histogram bin (HB) data for the ISN analysis, which are accumulated on board in 6° spin angle bins by the Combined Electronic Unit (CEU). Thus, the full spin contains 60 angular bins covering the entire great circle. HB data are used because direct event (DE) data suffered some undesirable losses due to the limited transmission capability of the interface and telemetry (Möbius et al. 2012, 2015a). HB data are accumulated over 64 IBEX spins, with each energy step observed for a total of 8 spins. For this study, we only use 6 bins with the spin angle centers between 252° and 282°. These bins have the highest count rates and their use minimizes the impact of other components (e.g., secondary helium) on the fitting of the parameters of the primary ISN helium (Swaczyna et al. 2015).

The interval selection (aka "good times") criteria applied to data in this study are as follows:

1. **Not spun times**. We use only periods for which the spin pulse is triggered correctly based on the orientation obtained from on-board attitude control system (ACS). An accurate spin pulse is necessary for proper accumulation of HB data. Sometimes, bright objects may blind the IBEX star tracker when they are present in its field of view. While the spin pulse is extrapolated based on the spin rate in such situations, the extrapolation error increases significantly over time. These periods are called "spun times." Effective correction procedures have been developed, but they can be applied only to DE data. We determine "spun times" based on star sensor measurements. The star sensor is aligned with IBEX-Lo and was intended to be a source of independent attitude data (Hłond et al. 2012). If the spin pulse is emitted correctly, the stars are visible at the same spin angle over the entire period. However, during "spun times", star positions drift. We manually select times for which these positions do not change.

2. **IBEX-Lo ENA good times**. The standard IBEX-Lo "good times" list (Galli et al. 2019, 2021) is used to further restrict the ISN data. This list excludes pixels and periods when the Moon or the Earth's magnetosphere is in the field of view. For orbits and arcs with the Oxygen or X-cal modes, this list is not created, and therefore, for these orbits, we do not use this criterion. However, we note that this criterion is typically less restrictive than the combination of requirements 1 and 3. We only use the times for which the bin range 10-20 is "good."

3. **Quiet TOF 2 times**. The last criterion selects only "quiet" times during which unwanted backgrounds are low. Following Leonard et al. (2015), we use the IBEX-Lo TOF 2 monitor rates collected by the instrument over wide 60° wide sectors. Since these rates are enhanced in the sector containing the core ISN flow, we use the rates from the two neighboring sectors as an added criterion. A detailed description of this criterion is provided in Appendix A.

We take periods that meet all three above criteria, except for the special mode, for which criterion (2) is not used. Together, the above criteria significantly restrict data available for the ISN study. In particular, "spun times" are common during the peak of the ISN season. Moreover, the last criterion eliminates long periods during the solar maximum in seasons 2014 and 2015. Nevertheless, with a complete solar cycle of ISN helium observations, the combined time coverage is much longer than that used in previous studies. Furthermore, we require that each good period cover an integer multiple of 512-spins, equivalent to approximately 2 hours of observations. These 512-spin blocks are the base periods reported in the derivative products posted concurrently with this paper (Appendix C).

### 2.3 Data Processing

IBEX data over the selected good times are accumulated for each orbit/arc. Each data point, identified here by $i$, correspond to a specific orbit/arc, ESA step, and spin angle bin. For each data point, we have the number of counts $d_i$ accumulated over the good times for this orbit and the total exposure time $t_i$. The raw count rate is given by the ratio of these two numbers.

The IBEX-Lo entrance system was designed to prevent ions and electrons from entering the instrument using the collimator bias voltage. Still a large number of electron events is detected by IBEX-Lo. While the electron events can be easily identified using the TOF subsystem, they were passed to the CEU causing throttling of the instrument interface buffer (Möbius et al. 2012). As a result, the electrons cause a significant load of TOF 3 only events, up to ~300-500 events per second in ESA step 2. With a maximum throughput of ~400-700 events per second, the probability of event loss, even with a double buffer used on IBEX-Lo, is significant. Moreover, since the event rate increases significantly in the ISN peak, the probability of data losses increases around the ISN peak. Swaczyna et al. (2015, Appendix) developed an algorithm estimating the total load of events through the interface buffer and thus the probability of event losses. We follow this algorithm to calculate correction factors $\gamma_i$, i.e., ratios of corrected count rates to the observed count rates, and their uncertainties $\delta\gamma_i$ for all orbits and arcs in the seasons affected by this issue. Starting with orbit 161, a new TOF logic was implemented, which requires valid TOF 2 measurements for events to be transmitted through the interface buffer. This change practically eliminated losses in the interface buffer starting with the 2013 ISN season.

Table 2
Characteristics of IBEX-Lo ISN Seasons

| Season | Total Counts | | | Exposure time | Orbits/arcs | Interface losses | PAC voltage | Spin offset |
|---|---|---|---|---|---|---|---|---|
| | ESA 1 | ESA 2 | ESA 3 | | | | | |
| 2009 | 37,871 | 43,877 | 43,378 | 5.37 hr | 5 orbits | Yes | 16 kV | - |
| 2010 | 33,509 | 38,374 | 40,848 | 4.09 hr | 4 orbits | Yes | 16 kV | - |
| 2011 | 0 | 81,981 | 0 | 2.89 hr | 4 orbits | Yes | 16 kV | - |
| 2012 | 17,646 | 64,513 | 20,001 | 4.95 hr | 6 arcs | Yes | 16 kV | - |
| 2013 | 30,762 | 40,186 | 48,662 | 10.59 hr | 10 arcs | No | 7 kV | - |
| 2014 | 24,375 | 31,661 | 39,017 | 7.45 hr | 8 arcs | No | 7 kV | - |
| 2015 | 5999 | 7961 | 9406 | 1.63 hr | 6 arcs | No | 7 kV | - |
| 2016 | 8457 | 11,029 | 0 | 1.53 hr | 3 arcs | No | 7 kV | - |
| 2017 | 63,825 | 98,028 | 55,113 | 11.55 hr | 8 arcs | No | 7 kV | 0.6° |
| 2018 | 73,224 | 94,107 | 57,157 | 16.49 hr | 8 arcs | No | 7 kV | 0.6° |
| 2019 | 108,485 | 137,243 | 92,287 | 18.81 hr | 8 arcs | No | 7 kV | 0.6° |
| 2020 | 105,292 | 133,862 | 88,637 | 16.91 hr | 8 arcs | No | 7 kV | 0.6° |

While the orbit/arc and angular bin selections include data dominated by the primary ISN helium, other contributions are subtracted before fitting the ISN flow parameters. A significant signal is expected from the Warm Breeze (Kubiak et al. 2014, 2016). The count rates (in s$^{-1}$) of the Warm Breeze contribution are estimated based on the Maxwellian approximation from Kubiak et al. (2016). The other contribution that we subtracted from the observed rates is related to the ubiquitous background as estimated by Galli et al. (2015, 2017) from the comparison of the corrected intensities for the ram and anti-ram observations of viewing directions that are not affected by magnetospheric contamination or ISN flows. The background rate is constant over periods of the same post-acceleration (PAC) voltage (see Section 3). For years 2009-2012, the background in ESA steps 1, 2, and 3 is 0.0098±0.0025 s$^{-1}$, 0.0089±0.0020 s$^{-1}$, and 0.0118±0.0015 s$^{-1}$, respectively. After the PAC voltage was lowered, these rates slightly dropped to 0.0067±0.0015 s$^{-1}$, 0.0075±0.0010 s$^{-1}$, and 0.0076±0.0018 s$^{-1}$, respectively.

Finally, count rate $c_i$ in bin $i$ for comparison with the simulations is calculated as follows:

$$c_i = \frac{\gamma_i d_i}{t_i} - w_i - b_i, \qquad (1)$$

where $d_i$ is the number of counts, $t_i$ is the bin accumulation time, $\gamma_i$ is the throughput correction factor, $w_i$ is the expected count rate contribution of the Warm Breeze, and $b_i$ is the ubiquitous background rate. The IBEX-Lo signal may contain additional sources of neutral atom fluxes, including ISN hydrogen atoms and heliospheric ENAs. These two sources are likely small compared to the primary ISN helium. ISN hydrogen atoms are mostly visible by IBEX during solar minima (Saul et al. 2012; Schwadron et al. 2013; Katushkina et al. 2015; Rahmanifard et al. 2019; Galli et al. 2019). Nevertheless, due to substantial uncertainties for the orbits with the core ISN helium flow, this contribution is smaller than the uncertainties obtained from their method. Still, Swaczyna et al. (2018) showed that the orbit/arc selection used here should minimize the impact of the ISN hydrogen contribution on fitting the primary ISN helium. The fluxes of low-energy ENAs, which may also contribute to the considered ESA steps, are small (Galli et al. 2016, 2017, 2021).

The IBEX-Lo subsystems and the entire instrument were calibrated before the launch (Fuselier et al. 2009). The calibration with low energy neutral atoms was challenging, and the obtained instrument geometric factor are not precise enough to be directly adopted in the study. Moreover, the change in the PAC voltage from 16 kV to 7 kV starting with orbit 177 decreased the instrument efficiency by approximately half. Galli et al. (2015) simulated the sputtered particle yield from the conversion surface and concluded that only helium atoms with energies higher than ~17 eV are observed with IBEX, in agreement with one of the conclusions from Sokół et al. (2015a). The energy range expected in the data analyzed here is significantly above this threshold (Swaczyna et al. 2018).

**Table 3**
**Uncertainty Sources in IBEX-Lo ISN Observations**

| Source | Parameter | Parameter unc. | Correlations |
|---|---|---|---|
| Statistical (Poisson) | $d$ | $\sqrt{d}$ | None |
| Throughput correction | $\gamma$ | see note | None |
| Background | $b$ | Section 2.3 | PAC&ESA |
| Warm Breeze – longitude | $\lambda_{WB}$ | 1.03° | All |
| Warm Breeze – latitude | $\beta_{WB}$ | 0.67° | All |
| Warm Breeze – speed | $v_{WB}$ | 0.85 km s$^{-1}$ | All |
| Warm Breeze – temperature | $T_{WB}$ | 1850 K | All |
| Warm Breeze – abundance | $n_{WB}$ | 0.004 | All |
| Warm Breeze – variation | $m_{WB}$ | 0.007 or 0.013 | ESA&Season |
| Spin axis R.A. | $\alpha$ | 0.01° | Orbit |
| Spin axis Dec. | $\delta$ | 0.01° | Orbit |
| Spin pulse accuracy | $\psi$ | 0.01° | Orbit |
| Boresight inclination | $\eta$ | 0.15° | All |
| Boresight spin angle | $\theta$ | 0.15° | All |
| Spin angle offset after 2016 | $\zeta$ | 0.07° | All>2016 |

The "good time" criteria discussed above return vastly different time coverages for the analyzed ISN seasons. Since statistical uncertainties are the main contributor to the final uncertainty, the statistical accuracy of each season depends predominantly on the number of observed counts. Table 2 presents the counts observed in the selected orbits/arcs, angular bins, and periods in each season, as well as the total exposure time of the selected angular bins. The table also shows the main factors that impacted the instrument performance over the 12 years of observations examined here. For nominal ESA stepping, counts expected in ESA 2 and 3 are comparable with each other, but in ESA 1 counts are slightly lower. However, these relations vary from season to season due to special modes and modified sweep tables. One may easily notice that seasons 2015 and 2016 include the lowest numbers of counts. In 2015, these low counts are due to the short length of good times, associated with the near-solar maximum conditions. In 2016, most data are not usable for ISN studies because of the star tracker anomaly. Thanks to the modified sweep table and relatively lengthy "good time" periods, the four most recent seasons have much higher counts that in the previous seasons. The spin offset reported in Table 2 is further discussed in Section 2.4 and Appendix B.

## *2.4 Data Uncertainty*

Swaczyna et al. (2015) introduced a detailed uncertainty analysis that includes correlations from various sources of uncertainties. While the statistical uncertainties are the main contributor to the overall uncertainty in most data points, other uncertainties are most important close to the ISN flow peak, where the count rates are high. These uncertainties also introduce correlations between data points, which need to be accounted for. In this study, we follow the uncertainty analysis performed by Swaczyna et al. (2015). A list of uncertainty sources included in this analysis is provided in Table 3.

If correlations between factors contributing to the final uncertainty are neglected, then each uncertainty component is described as a separate covariance matrix ($\mathbf{V}^\epsilon$). The complete covariance matrix is given as a sum of these matrices: $\mathbf{V} = \sum_\epsilon \mathbf{V}^\epsilon$. Element $(i,j)$ in each component matrix can be generally written in the following form:

$$\mathbf{V}_{ij}^{\epsilon} = \frac{\partial c_i}{\partial \epsilon} \frac{\partial c_j}{\partial \epsilon} (\delta\epsilon)^2, \qquad (2)$$

where $\frac{\partial c_i}{\partial \epsilon}$ is the partial derivative of the corrected count rate $c_i$ with respect to parameter $\epsilon$, and $\delta\epsilon$ denotes the uncertainty of this parameter. With this formulation, many entries of such matrices are equal to zero. For example, the count rate in a data point $i$ does not depend on the number of counts $d_j$ for $j \neq i$, and thus only one entry for $\epsilon = d_i$ is nonzero. Consequently, all statistical uncertainties are combined to a single covariance matrix in the following form:

$$\mathbf{V}_{ij}^{d} = \left(\frac{\partial c_i}{\partial d_i}\right)^2 (\delta d_i)^2 \delta_{i,j} = \left(\frac{\gamma_i}{t_i}\sqrt{d_i}\right)^2 \delta_{i,j}. \qquad (3)$$

In this equation, we use the relation between count rates and counts given by Equation (1), and we use the approximation that the Poisson uncertainty of counts is $\delta d_i = \sqrt{d_i}$. The Kronecker delta $\delta_{i,j}$ denotes that the statistical uncertainties are not correlated, and thus only diagonal elements are nonzero. A more general form can be applied to all uncertainty sources listed in Table 3:

$$\mathbf{V}_{ij}^{\epsilon} = \delta_\epsilon c_i \delta_\epsilon c_j \delta_{f(i),f(j)}, \qquad (4)$$

where $\delta_\epsilon c_i = \frac{\partial c_i}{\partial \epsilon}\delta\epsilon$. The Kronecker delta here shows where the same parameter describing the uncertainty is applied to more than one data point. The function $f(i)$ is shown for each uncertainty in column Correlations in Table 3. If two characteristics are mentioned in this column, the Kronecker delta is 1 only for points that have the same value of both characteristics. The correlation "All" denotes that a matrix of ones should be used in place of the Kronecker delta in Equation (4). Most elements of the IBEX-Lo ISN uncertainty matrix have been extensively discussed by Swaczyna et al. (2015). However, we introduce here three new uncertainty sources.

The first new uncertainty source is related to the possible variation of the Warm Breeze signal between years and in the analyzed ESA steps. Kubiak et al. (2016) used only ESA 2 to find the Warm Breeze. Swaczyna et al. (2018) noted that the Warm Breeze shows a stronger variation in signal strength between ESA steps than the variation in primary ISN helium. Consequently, we add an uncertainty of 20% to the abundance of Warm Breeze in ESA 1 and 3. This value is established by inspection of the variation between ESA steps for orbits and bins dominated by the Warm Breeze. Additionally, we also include yearly variations in this new component ($m_{\mathrm{WB}}$ in Table 3). In summary, the relative uncertainty is $\delta m_{\mathrm{WB}} = 0.007$ for ESA 2 and $\delta m_{\mathrm{WB}} = 0.013$ for ESA 1 and 3. This uncertainty correlates data points with the same ESA step and observation seasons.

Another uncertainty source that was omitted in the previous study is related to the accuracy of the spin pulse, which begins the accumulation of histograms for each spin. This accuracy corresponds to the average position of the spin pulse for a given orbit. We estimate this uncertainty to be of the same order as the uncertainty of the spin axis determination. The derivative with respect to the spin angle is calculated here using simulations with shifted spin pulse positions. This method also applies here to the unknown position of the boresight direction in spin angle.

The last uncertainty listed in Table 3 is related to the spin angle offset that was inadvertently introduced starting in orbit 326 when a software change in the spacecraft ACS was included. This software change was necessary to restore operation of the star tracker after an anomaly starting during orbit 315b. Our initial results of the ISN helium analysis showed that the ecliptic latitude of the primary ISN helium flow was systematically shifted by ~0.5° between the observations collected in the seasons 2009-2016 and 2017-2020. Since a shift in the ecliptic latitude is directly related to a shift in the spin angle, we surmised that this change may have been caused by a systematic shift in the spin pulse. Fortunately, thanks to the star sensor installed on IBEX-Lo, which was intended to support the determination of the spacecraft orientation (Hłond et al. 2012), we verify this shift using an independent dataset and determine its magnitude. As shown in Appendix B, we find a shift of 0.60°±0.07° if we compare star sensor observations before and after the change in the ACS software. We account for this shift in the simulations, and the uncertainty of this shift is included in our uncertainty system.

### 3. Simulations of IBEX Signal and Normalization Factors

The ISN helium parameters are determined in this study by $\chi^2$ minimization of the observed count rates compared to the expected fluxes modeled using the numerical version of the Warsaw Test Particle Model (WTPM, Sokół et al. 2015b). The numerical WTPM, differently than the analytic version, utilizes a time-dependent ionization model varying along ISN atom trajectories. The WTPM integrates neutral atom trajectories in the heliosphere from the source region at 150 au from the Sun (i.e., beyond the heliopause) to IBEX at 1 au. The distribution function in the source region is assumed Maxwellian, characterized by the temperature and bulk velocity relative to the Sun at 150 au from the Sun. Therefore, the modeled ISN helium fluxes depend on four parameters: the flow speed, temperature, and inflow direction (ecliptic longitude and latitude).

Helium atoms are ionized on their journey from the VLISM to IBEX. The WTPM includes time- and heliolatitude-dependent models of the ionization processes. Helium atoms are ionized through photoionization, charge exchange with the solar wind ions, and electron impact ionization (Bzowski et al. 2013; Sokół & Bzowski 2014; Sokół et al. 2019a, 2020). Photoionization is the dominant process and shows a significant variation during the solar activity cycle. The photoionization rate in the WTPM is adopted from Sokół et al. (2020). Electron-impact ionization strongly fluctuates on time scales shorter than one Carrington rotation (the nominal time resolution of the model). However, during the 12-year interval, its contribution to the total rate at 1 au was only between 10% during high solar activity and 15% during low activity (Sokół et al. 2020). Furthermore, the rate of this reaction falls off with the solar distance (Rucinski & Fahr 1989; Bzowski et al. 2013) more rapidly than the photoionization rate and becomes negligible outside of ~2 au. Charge exchange with solar wind protons is also almost negligible because of a very low cross section for this reaction (Phaneuf et al. 1987), and the dominant charge exchange process for ISN He is that with solar wind alpha particles. Overall, charge exchange contributes between 1% and 5% to the total ionization rate (Sokół et al. 2020).

The distribution function of the ISN helium at IBEX calculated using the WTPM is integrated over the speed, angular transmission function of the instrument collimator, spin angle, and good time intervals for the positions of the strip of the sky visible to the instrument corresponding to the orientation of the IBEX spin axis for a given IBEX orbit/arc. The details of these integrations are described in Sokół et al. (2015b). The result of this integration is the average flux of ISN helium atoms (in cm$^{-2}$ s$^{-1}$ sr$^{-1}$) entering the instrument within a given spin-angle bin averaged over the exposure period. Therefore, these simulations correspond to the specific observational conditions, which account for spacecraft orientation, position, and motion relative to the Sun. To include the energetic response of IBEX-Lo, the model additionally provides the first (i.e., the mean speed) and the second non-central moments of ISN helium atoms speeds in each bin (Swaczyna et al. 2018).

This study employs a new fitting procedure that utilizes polynomial interpolation with coefficients obtained from finite differences (see details in Section 4). For this procedure, we calculated ISN helium fluxes for a baseline parameter set and parameter sets in which one or two parameters are modified by quantized variations. As the baseline inflow parameter set, we used the result reported by Swaczyna et al. (2018, Line 9 in Table 2), i.e., $\boldsymbol{\pi}^0 = (\lambda^0, \beta^0, v^0, T^0) = (255.62°, 5.16°, 25.82 \text{ km s}^{-1}, 7673 \text{ K})$, which represent the ecliptic longitude, latitude, speed, and temperature, respectively. As the base quantized variations, we use 1/3 of the maximum deviation between the above baseline set and those reported by Swaczyna et al. (2018, Table 2) or Bzowski et al. (2015, Table 2). These quantized variations are $\Delta\lambda = 1.70°$, $\Delta\beta = 0.08°$, $\Delta v = 1.21 \text{ km s}^{-1}$, $\Delta T = 804$ K. Variations are added or subtracted from the baseline values, and the WTPM is used to calculate the expected

## Table 4
## Scaling Function Forms

| Name | Form | Scaling Function Coefficients $q$ | #Coeff |
|---|---|---|---|
| Y-norm | $S_i = a_{\text{ESA,year}}$ | $a_{1,2009}, a_{1,2010}, \ldots, a_{1,2020}, a_{2,2009}, \ldots, a_{2,2020}, a_{3,2009}, \ldots, a_{3,2020}$ | 33 |
| YV-norm | $S_i = a_{\text{ESA,year}}[1 + b_{\text{ESA,PAC}}(v_i - v_0)]$ | $a_{1,2009}, a_{1,2010}, \ldots, a_{3,2020}, b_{1,\text{N}}, b_{2,\text{N}}, b_{3,\text{N}}, b_{1,\text{L}}, b_{2,\text{L}}, b_{3,\text{L}}$ | 39 |
| R-norm | $S_i = r_{\text{ESA,PAC}} a_{\text{year}}$ | $r_{1,\text{N}}, r_{3,\text{N}}, r_{1,\text{L}}, r_{3,\text{L}}, a_{2009}, a_{2010}, \ldots, a_{2020}$ | 16 |
| RV-norm | $S_i = r_{\text{ESA,PAC}} a_{\text{year}}[1 + b_{\text{ESA,PAC}}(v_i - v_0)]$ | $r_{1,\text{N}}, r_{3,\text{N}}, r_{1,\text{L}}, r_{3,\text{L}}, a_{2009}, a_{2010}, \ldots, a_{2020}, b_{1,\text{N}}, b_{2,\text{N}}, b_{3,\text{N}}, b_{1,\text{L}}, b_{2,\text{L}}, b_{3,\text{L}}$ | 22 |
| P-norm | $S_i = a_{\text{ESA,PAC}}$ | $a_{1,\text{N}}, a_{2,\text{N}}, a_{3,\text{N}}, a_{1,\text{L}}, a_{2,\text{L}}, a_{3,\text{L}}$ | 6 |
| PV-norm | $S_i = a_{\text{ESA,PAC}}[1 + b_{\text{ESA,PAC}}(v_i - v_0)]$ | $a_{1,\text{N}}, a_{2,\text{N}}, a_{3,\text{N}}, a_{1,\text{L}}, a_{2,\text{L}}, a_{3,\text{L}}, b_{1,\text{N}}, b_{2,\text{N}}, b_{3,\text{N}}, b_{1,\text{L}}, b_{2,\text{L}}, b_{3,\text{L}}$ | 12 |

fluxes for these parameter combinations. With this arrangement, we are able to perform second-degree multivariate Newton polynomial interpolation that also accounts for mixed second derivative terms.

Following the convention established in our previous studies (Bzowski et al. 2015; Sokół et al. 2015b; Swaczyna et al. 2015, 2018), the results of the model for a bin $i$ (orbit/arc and angular bin) are denoted as $F_i(\boldsymbol{\pi})$, where $\boldsymbol{\pi} = (\lambda, \beta, v, T)$ is a vector representing the physical parameters of the flow. Each bin is typically characterized by the orbit/arc, ESA step, and spin angle bin. The mean velocity calculated from the model is denoted as $v_i(\boldsymbol{\pi})$.

Following Swaczyna et al. (2018), we apply scaling functions (norms) to the simulated flux to model the instrument's energetic response function. The scaling functions represent the effective geometric factor of the instrument (in cm² sr). Furthermore, the range of ISN helium atom speeds that we consider in this study is relatively narrow (see Figure 3 in Swaczyna et al. 2018). Therefore, we model this response either as a linear function of the mean atom speed in the spacecraft frame, or as a constant factor. The forms of the scaling function adopted in this study are listed in Table 4.

Each of the considered forms have several coefficients, listed in Table 4. The coefficients are mostly separate for each ESA step, and we expect that they may have changed when the PAC voltage was decreased from 16 kV (nominal – N) to 7 kV (low – L). For the P- and PV-norms, only PAC voltage and ESA step characterize the coefficients. Swaczyna et al. (2018) concluded that these norms were sufficient to characterize some temporal changes of the instrument performance. This approach, however, tacitly precluded effects of possible changes of the instrument sensitivity with time or the possible existence of systematic changes in the ISN He flux missed by the model. Therefore, we also use the Y- and YV-norms, allowing the proportionality coefficient to vary from year to year, which is the form used in some previous studies (Bzowski et al. 2012, 2015; Swaczyna et al. 2015). Finally, we introduce the R- and RV-norms here in which the efficiency can change from year to year, but the ratio between the ESA steps may only differ for the nominal and low PAC voltages. Note that we use ESA 2 as a basis of this ratio, i.e., $r_{2,\text{N}} = r_{2,\text{L}} = 1$.

Velocity-dependent scaling functions (*V-norms) have an additional factor $[1 + b_{\text{ESA,PAC}}(v_i - v_0)]$, which is the same for all considered norms. The coefficient in this factor depends on the ESA step and the PAC voltage (nominal or low). Following Swaczyna et al. (2018), we select a reference speed of $v_0 = 78$ km s$^{-1}$. At this speed, this factor is equal 1, thus the linear coefficients represent the geometric factor at this speed. The vector of the coefficients describing the scaling function in each case is denoted later as $\boldsymbol{q}$. The number of coefficients for each form is listed in the last column of Table 4.

## 4. Fitting Simulations to Data

Parameter fitting is performed by minimization of the $\chi^2$-statistic in the form (Swaczyna et al. 2015, 2018):

$$\chi^2(\boldsymbol{\pi}, \boldsymbol{q}) = \sum_{i,j} [c_i - g_i(\boldsymbol{\pi}, \boldsymbol{q})] \, (\mathbf{V}^{-1})_{i,j} [c_j - g_j(\boldsymbol{\pi}, \boldsymbol{q})], \tag{5}$$

where $c_i$ is the corrected count rate in a bin $i$ obtained from Equation (1), $g_i(\boldsymbol{\pi}, \boldsymbol{q}) = S_i(\boldsymbol{\pi}, \boldsymbol{q})F_i(\boldsymbol{\pi})$ is a product of the scaling function given in Table 4 and the modeled flux $F_i(\boldsymbol{\pi})$, and $(\mathbf{V}^{-1})_{i,j}$ represent element $(i,j)$ of the inverted uncertainty matrix (see Section 2.4). The scaling function coefficients $\boldsymbol{q}$ are nuisance parameters, which from the perspective of finding the best-fit parameters are not distinguishable from the sought ISN helium parameters $\boldsymbol{\pi}$. Therefore, we perform the minimization with respect to all parameters $(\boldsymbol{\pi}, \boldsymbol{q})$.

The product $g_i(\boldsymbol{\pi}, \boldsymbol{q}) = S_i(\boldsymbol{\pi}, \boldsymbol{q})F_i(\boldsymbol{\pi})$ depends on the scaling parameters analytically as expressed in Table 4. However, the mean ISN helium atom speed $v_i(\boldsymbol{\pi})$ and flux $F_i(\boldsymbol{\pi})$ are calculated numerically for specified parameters of ISN helium. Therefore, we use a second-degree multivariate Newton polynomial to interpolate (for simplicity the scaling parameter coefficients $\boldsymbol{q}$ are omitted in the equations below, since they are not part of the interpolation):

$$g_i(\boldsymbol{\pi}) = g_i(\boldsymbol{\pi}^0) + \sum_{p=1}^{4} (\pi_p - \pi_p^0)\Delta_p[g_i(\boldsymbol{\pi}^0)] + \frac{1}{2}\sum_{p=1}^{4}\sum_{r=1}^{4} (\pi_p - \pi_p^0)(\pi_r - \pi_r^0)\Delta_{p,r}^2[g_i(\boldsymbol{\pi}^0)], \quad (6)$$

where $\pi_p$ and $\pi_p^0$ denote $p$th element vectors $\boldsymbol{\pi}$ and $\boldsymbol{\pi}^0$. The finite differences are calculated from the following formulas:

$$\Delta_p[g_i(\boldsymbol{\pi}^0)] = \frac{g_i(\boldsymbol{\pi}^0 + \boldsymbol{\rho}^p) - g_i(\boldsymbol{\pi}^0 - \boldsymbol{\rho}^p)}{2|\boldsymbol{\rho}^p|}, \quad (7)$$

$$\Delta_{p,p}^2[g_i(\boldsymbol{\pi}^0)] = \frac{g_i(\boldsymbol{\pi}^0 + \boldsymbol{\rho}^p) - 2g_i(\boldsymbol{\pi}^0) + g_i(\boldsymbol{\pi}^0 - \boldsymbol{\rho}^p)}{|\boldsymbol{\rho}^p|^2}, \quad (8)$$

$$\Delta_{p,r}^2[g_i(\boldsymbol{\pi}^0)] = \frac{g_i(\boldsymbol{\pi}^0 + \boldsymbol{\rho}^p + \boldsymbol{\rho}^r) - g_i(\boldsymbol{\pi}^0 + \boldsymbol{\rho}^p - \boldsymbol{\rho}^r) - g_i(\boldsymbol{\pi}^0 - \boldsymbol{\rho}^p + \boldsymbol{\rho}^r) + g_i(\boldsymbol{\pi}^0 - \boldsymbol{\rho}^p - \boldsymbol{\rho}^r)}{4|\boldsymbol{\rho}^p||\boldsymbol{\rho}^r|}, \quad (9)$$

where the vector $\boldsymbol{\rho}^p$ is defined as:

$$\boldsymbol{\rho}^p = (\delta_{1,p}\Delta\lambda, \delta_{2,p}\Delta\beta, \delta_{3,p}\Delta v, \delta_{4,p}\Delta T), \quad (10)$$

i.e., it denotes a parameter shift in the $p$th parameter with its quantized variation. The fluxes and the mean speeds of ISN helium atoms in all considered directions of this shift are calculated as discussed in Section 3. Based on this approximation, the formula given in Equation (6) is a second-degree multivariate polynomial in the inflow parameters $(\lambda, \beta, v, T)$. This polynomial in the analytic form is directly applied to Equation (5). The polynomial obtained from this equation is a higher degree multivariate polynomial mixing the inflow parameters and scaling function coefficients. We find numerically the minimum of the $\chi^2$-statistic with respect to all these parameters and coefficients. This method is both efficient and accurate in finding the minimum. We verify that the same minimum is obtained with two sets of starting parameters, the first one in which the initial parameters are equal to the baseline parameter set, and the other one in which we add a random variation with a magnitude given by quantized variations discussed in Section 3. The set $(\boldsymbol{\pi}^m, \boldsymbol{q}^m)$ for which the statistic is minimal are called the best-fit parameters.

The interpolation method introduced in this study differs from the approach proposed by Swaczyna et al. (2015). They calculated the $\chi^2$-statistic for several hundred sets of interstellar flow parameters and interpolated this statistic in the proximity of the parameter set for which the statistics was the smallest. Unfortunately, that methodology causes problems as the parameter correlation tube narrows, and thus the minimum over a limited number of parameters may be significantly shifted from the actual minimum.

While the $\chi^2$ statistics provide the best fit parameters for each considered form of the scaling functions as discussed in Section 3, it does not directly allow for a comparison of models with different numbers of fit parameters. While we have 4 physical parameters $\boldsymbol{\pi}$ of the flow in all cases, there is a variable number of parameters $\boldsymbol{q}$ describing the scaling function as listed in Table 4. To select between these models, we use two criteria: the Akaike Information Criterion (AIC, Akaike 1974) and the Bayesian Information Criterion (BIC, Schwarz 1978), which are calculated from the following relations:

$$\text{AIC} = \chi^2 + 2K, \tag{11}$$

$$\text{BIC} = \chi^2 + K \log N, \tag{12}$$

where $K$ is the number of model parameters, and $N$ is the number of data points. Both criteria add a penalty term proportional to the number of model parameters, but with different proportionality factors. The model with lower values of these criteria should be considered the better model.

Uncertainties of the model parameters as a covariance matrix are obtained from second derivatives of the $\chi^2$ statistics at the minimum (Particle Data Group et al. 2020):

$$\boldsymbol{\Sigma} = \left( \frac{1}{2} \frac{\partial \chi^2}{\partial (\boldsymbol{\pi}, \boldsymbol{q})^2} \right)^{-1}. \tag{13}$$

The related correlation matrix can be obtained by dividing each matrix entry by the square roots of diagonal elements from the associated row and column. Finally, the covariance matrix is multiplied by the reduced $\chi^2$, i.e., the values of the statistics $\chi^2$ at the minimum divided by the number of degrees of freedom $\nu = N - K$. This multiplication is only applied if the reduced $\chi^2$ is above 1. Such a situation indicates that the model does not fully describe the data or that the uncertainty analysis is incomplete. This scaling is equivalent to scaling of the data uncertainty matrix by a constant factor. All reported uncertainties and plots in this paper account for this scaling factor.

## 5. Results: Best-Fit Interstellar Flow Parameters

The $\chi^2$ minimization, as described in Section 4, has been performed for the full dataset from seasons 2009-2020, ESA steps 1, 2, and 3, and the orbit and bin selection as discussed in Section 2. The results for various forms of the scaling function, as listed in Table 4, are presented in Table 5. The table summarizes the fit goodness criteria and lists the ISN helium flow parameters for each of the considered forms of the scaling function. The preferred scaling functions based on AIC and BIC are YV-norm and RV-norm, respectively. Since all of these functions serve as an effective description of IBEX-Lo energy response, not a physical model of the response function, BIC is the preferred criterion to consider in such situations (Kadane & Lazar 2004). On the other hand, AIC is preferred where different physical models with different number of parameters are compared. Nevertheless, we notice that the flow parameters are almost identical, i.e., differences for these two cases are about 10 times smaller than the uncertainty in each considered ISN helium flow parameter. Therefore, we choose to use the results obtained with RV-norm in this study.

**Table 5**
**Results of the fitting to the entire dataset**

| Norm | $N$ | $K$ | $\nu$ | $\chi^2$ | $\chi^2/\nu$ | AIC | BIC | $\lambda$ (°) | $\beta$ (°) | $v$ (km s$^{-1}$) | $T$ (10$^3$ K) |
|---|---|---|---|---|---|---|---|---|---|---|---|
| Y-norm | 1254 | 37 | 1217 | 2085.6 | 1.714 | 2159.6 | 2349.6 | 255.88±0.23 | 5.22±0.09 | 25.64±0.22 | 7.42±0.14 |
| YV-norm | 1254 | 43 | 1211 | 1818.6 | 1.502 | 1904.6 | 2125.4 | 255.59±0.23 | 5.13±0.08 | 25.84±0.21 | 7.43±0.14 |
| R-norm | 1254 | 20 | 1234 | 2137.4 | 1.732 | 2177.4 | 2280.1 | 255.89±0.23 | 5.22±0.10 | 25.67±0.22 | 7.43±0.14 |
| RV-norm | 1254 | 26 | 1228 | 1858.7 | 1.514 | 1910.7 | 2044.2 | 255.59±0.23 | 5.14±0.08 | 25.86±0.21 | 7.45±0.14 |
| P-norm | 1254 | 10 | 1244 | 6406.8 | 5.150 | 6426.8 | 6478.1 | 256.45±0.37 | 5.08±0.17 | 25.42±0.31 | 7.18±0.20 |
| PV-norm | 1254 | 16 | 1238 | 6154.1 | 4.971 | 6186.1 | 6268.2 | 257.10±0.41 | 5.17±0.17 | 24.77±0.35 | 6.88±0.21 |

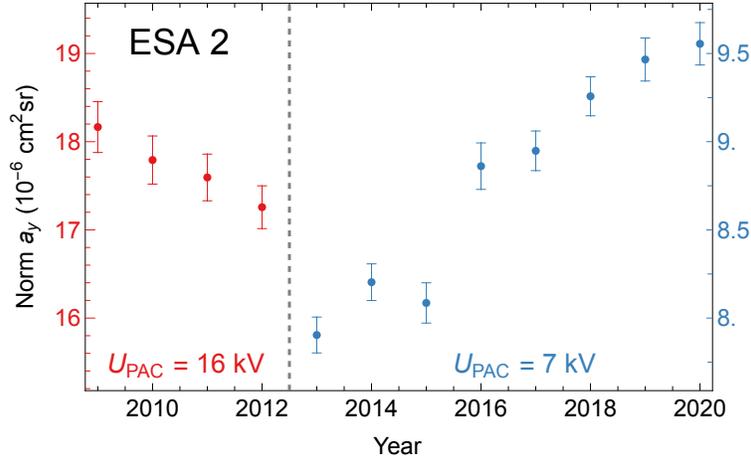

**Figure 1** – Linear coefficients of the RV-norm scaling function for each observational season. The red and blue symbols correspond to periods with the nominal and the lowered PAC voltage, respectively, with the coefficient value provided on the left and right axis respectively.

Swaczyna et al. (2018) concluded that the scaling functions that depend only on PAC voltage (P- and PV-norm) can be used to model the instrument response instead of annually changing normalization factors. They noticed that the efficiency drops about two-fold when the PAC voltage is decreased. Our new results are in apparent disagreement with this finding since the reduced $\chi^2$ increases more than 3-fold if such norms are applied. However, in the previous study, only the first seven observational seasons were available. Figure 1 presents the best fit ESA 2 norms obtained with the RV-norm scaling function. This plot shows that the coefficients are marginally consistent with single values for the periods 2009-2012 and 2013-2015. These two periods were previously assigned to the nominal, and the lowered PAC voltage, respectively. In other words, the apparent agreement of these coefficients in the previous study appears to be coincidental. The time variation of this coefficient is further discussed in Section 6.

With the RV-norms, ratios between the scaling function for ESA 1 and 3 to ESA 2 are $r_{1,N} = 0.806 \pm 0.012$, and $r_{3,N} = 1.110 \pm 0.017$ under the nominal 16 kV PAC voltage, and $r_{1,L} = 0.758 \pm 0.005$, and $r_{3,L} = 1.173 \pm 0.008$ under the low 7 kV PAC voltage. These ratios demonstrate that after the PAC voltage was lowered, the relative efficiency of ESA 1 decreased, while the relative efficiency of ESA 3 increased compared to ESA 2. Such changes are expected since the detection efficiency of less energetic atoms increases with increasing PAC voltage. Nevertheless, these changes are small compared to a ~2-fold absolute efficiency decrease, as shown in Figure 1. The best-fit coefficients describing the speed-dependence of the scaling function are: $b_{1,N} = -0.021 \pm 0.007$, $b_{2,N} = -0.037 \pm 0.006$, $b_{3,N} = 0.004 \pm 0.007$ for the nominal, 16 kV, PAC voltage, and $b_{1,L} = -0.020 \pm 0.005$, $b_{2,L} = -0.041 \pm 0.004$, $b_{3,N} = -0.003 \pm 0.005$ for the lower, 7 kV, PAC voltage. The speed-dependent coefficient is statistically consistent with 0 for ESA 3, in agreement with the assumption used for the in-flight calibration (Schwadron et al. 2022). The coefficients for the respective ESA steps before and after the change in the PAC voltage remained almost unchanged.

The minimum value of $\chi^2$ is approximately 50% larger than that statistically expected for the number of degrees of freedom of the fit. We scale the final uncertainty as described in Section 4 to compensate for this discrepancy.

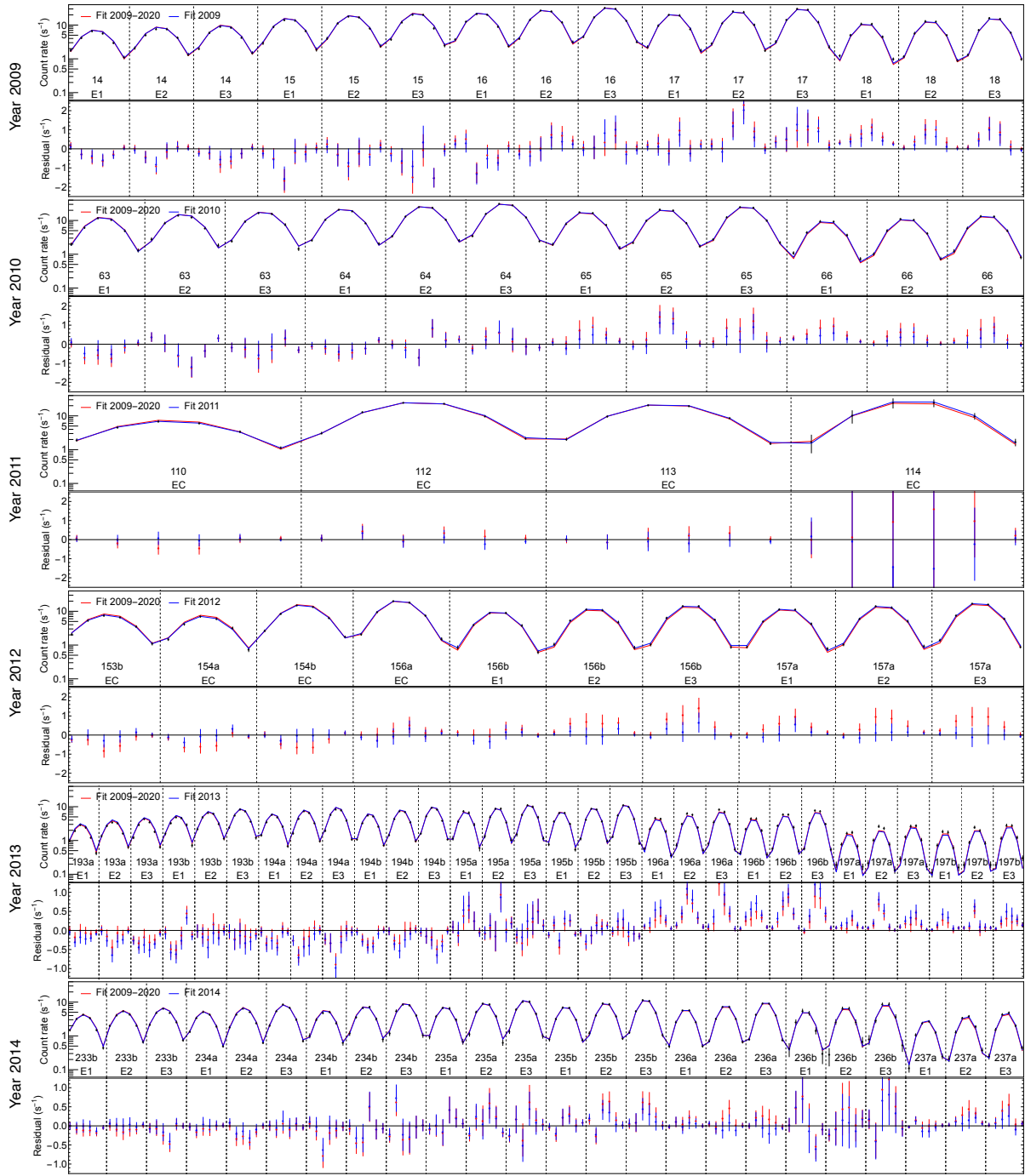

**Figure 2** – Comparison of the best-fit model using the RV-norm with the IBEX data. Plots from top to bottom correspond to observational seasons 2009-2014. The top part of each panel compares the data with the models, and the bottom part shows the residual signal. Red and blue symbols show results for the global fit and the fit with this observational season, respectively.

However, this discrepancy may result from signal sources in IBEX data that are not accounted for. As discussed by Swaczyna et al. (2018), ISN hydrogen atoms may increase the observed count rates in each observing season, especially for later orbits. Still, the reduced data range used in this study should result in relatively small

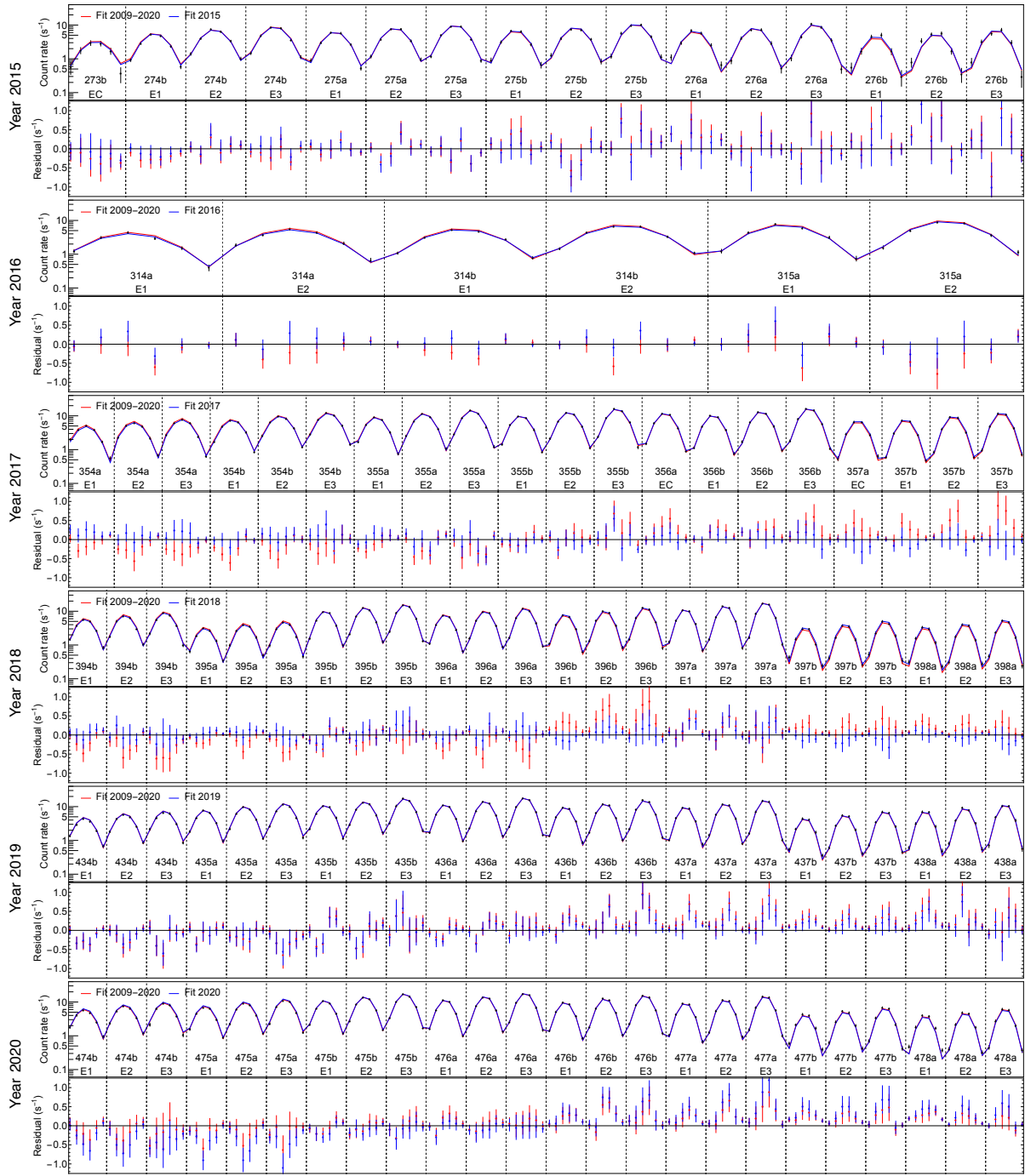

**Figure 2** (*cont.*) – Results for observational seasons 2015-2020 (top to bottom).

contributions of these atoms. Figure 2 shows a comparison of the IBEX data with the global best-fit model and the best-fit to each year separately (see Section 7). This comparison shows that the residuals are not randomly scattered. Rather the residuals often show systematic surplus or deficit of observed flux for each orbit. Moreover, the residuals show a systematic pattern with negative residuals early in each observational season before the peak and positive residuals after the orbit with the peak. This pattern suggests that the hypothetical contribution from the ISN hydrogen is not eliminated and requires further analysis, beyond the scope of this paper. The model

**Table 6**
**Comparison of the ISN He Flow Parameters from Select Studies**

| Source | Instrument | Years | $\chi^2/\nu$ | $\lambda$ (°) | $\beta$ (°) | $v$ (km s$^{-1}$) | $T$ (10$^3$ K) |
|---|---|---|---|---|---|---|---|
| Bzowski et al. (2014) | Ulysses/GAS | 1994-2007 | 2.16 | 255.3 | 6.0 | 26.0 | 7.47 |
| Wood et al. (2015) | Ulysses/GAS | 1994-2007 | 1.52 | 255.54±0.19 | 5.44±0.24 | 26.08±0.21 | 7.26±0.27 |
| Bzowski et al. (2015) | IBEX-Lo | 2009-2014 | 1.84 | 255.8±0.5 | 5.16±0.10 | 25.8±0.4 | 7.44±0.26 |
| Swaczyna et al. (2018 P-norm) | IBEX-Lo | 2009-2015 | 1.64 | 255.62±0.36 | 5.16±0.08 | 25.82±0.33 | 7.67±0.23 |
| Swaczyna et al. (2018 PV-norm) | IBEX-Lo | 2009-2015 | 1.49 | 255.41±0.40 | 5.03±0.07 | 26.21±0.37 | 7.69±0.23 |
| Taut et al. (2018) | STEREO A /PLASTIC | 2007-2014 | … | 255.41±0.34 | … | … | … |
| Wood et al. (2019)[a] | IBEX-Lo | 2011-2014 | 1.83 | 255.6±1.5 | 5.09±0.14 | 25.6±1.2 | $T_\perp = 7.58\pm0.96$ $T_\parallel = 12.70\pm2.96$ |
| This study (RV-norm) | IBEX-Lo | 2009-2020 | 1.51 | 255.59±0.23 | 5.14±0.08 | 25.86±0.21 | 7.45±0.14 |

**Notes.** [a] Bi-Maxwellian distribution with symmetry axis pointing ecliptic $(\lambda_{\text{axis}}, \beta_{\text{axis}}) = (57.2° \pm 8.9°, -1.6° \pm 5.9°)$.

fits with individual years marked with the blue line in each plot shows a slightly smaller residual flux. The smaller residuals are expected since the parameters may be adjusted from the best global fit to better follow observations from each year. Further discussion of results obtained from individual years is provided in Section 7.

While the scaling function parameters describe the instrument response, we are mostly interested in the parameters describing the primary ISN helium population in the VLISM near the heliosphere. Table 6 compares the final parameters obtained in this study with previous results. Our current best-fit parameters are almost identical with the results from Bzowski et al. (2015) and Swaczyna et al. (2018, P-norm), which were obtained assuming that the scaling function is not speed-dependent. However, the parameters for the velocity-dependent scaling function (PV-norm) from Swaczyna et al. (2018) differ significantly from the current result. The apparent consistency with the speed-independent norms is likely a coincidence. The previous study used a subset of the dataset of IBEX observations used here, resulting in a much smaller number of total counts (see Table 2). The inflow longitude is also consistent with the result obtained from observations of pickup ion cutoff by PLASTIC on STEREO A (Möbius et al. 2015b, 2016; Taut et al. 2018; Bower et al. 2019).

The observational strategy of IBEX results in a significant correlation between the inflow parameters (Lee et al. 2012, 2015; McComas et al. 2012; Möbius et al. 2015a; Schwadron et al. 2022). Because of this correlation, the uncertainties should be reported together with the correlation matrix between the inflow parameters. The correlation matrix between the flow parameters for RV-norm is:

$$\text{Cor} = \begin{pmatrix} 1 & -0.279 & -0.844 & -0.861 \\ -0.279 & 1 & 0.316 & 0.364 \\ -0.844 & 0.316 & 1 & 0.962 \\ -0.861 & 0.364 & 0.962 & 1 \end{pmatrix}, \tag{14}$$

where the column and row orders correspond to the inflow longitude, latitude, speed, and temperature. The largest correlation (0.962) is between the speed and temperature, which is not surprising, as IBEX measures the Mach cone angle of the interstellar flow with very high accuracy. There is also a strong anti-correlation of the inflow longitude with the speed (–0.844) and the temperature (–0.861).

This analysis provides the ISN helium parameters at the heliopause (150 au), which differ from the pristine VLISM conditions. Two effects modify the inflow speed, direction, and temperature between the pristine VLISM and the heliopause, as seen from the IBEX vantage point. First, solar gravity accelerates and slightly deflects ISN atoms before they enter the heliosphere. McComas et al. (2015b) compared best-fit parameters obtained with the WTPM that tracked atoms to 150 au and 1000 au. Over this distance, the apparent bulk speed of the population increases by $\Delta v_{\text{grav}} = 0.4$ km s$^{-1}$, the inflow ecliptic longitude decreases by $\Delta \lambda_{\text{grav}} = -0.1°$, while the temperature increases by $\Delta T_{\text{grav}} = 200$ K. Since IBEX observes the broad beam of ISN helium, the modifications to the flow velocity from the comparison shown in McComas et al. (2015b) are smaller than gravitational deflection of a cold ISN helium atom detected by IBEX in the atom orbit perihelion. Additionally, Swaczyna et al. (2021) showed that the ISN helium atoms are heated and slowed down by elastic collisions in the outer heliosheath. These collisions decrease the bulk speed by $\Delta v_{\text{el.coll.}} = -0.45$ km s$^{-1}$, and increase the temperature by $\Delta T_{\text{el.coll.}} = 1100$ K. Consequently, the best estimates of the pristine VLISM parameters from the current study accounting for these two effects are $v_{\text{VLISM}} = v_{\text{HP}} - \Delta v_{\text{grav}} - \Delta v_{\text{el.coll.}} = 25.9$ km s$^{-1}$, and $T_{\text{VLISM}} = T_{\text{HP}} - \Delta T_{\text{grav}} - \Delta T_{\text{el.coll.}} = 6150$ K.

## 6. Temporal Variability of Scaling Function

The best-fit linear coefficients of the scaling function for the RV-norm, as shown in Figure 1, suggest that the instrument efficiency changes over time. Figure 1 indicates that the efficiency of IBEX-Lo decreased in the first four years of the mission but increased over the later period. While the decrease might be associated with some sort of reduction in detection efficiency, the increase between 2013 and 2020 is difficult to explain.

The average ratio of the RV-norm at the nominal speed between the lowered and the nominal PAC voltage given as $\xi_{\text{ESA}} = \frac{r_{\text{ESA,L}}}{r_{\text{ESA,N}}} \frac{\langle a_y \rangle_{y=2013,\ldots 2020}}{\langle a_y \rangle_{y=2009,\ldots 2012}}$ are 0.467±0.009, 0.496±0.004, and 0.524±0.010 for ESA 1, 2, and 3, respectively. During the IBEX-Lo post-calibration campaign, the reduced TOF detection efficiency has been measured in the laboratory to be 0.434±0.021, 0.435±0.021, and 0.437±0.022 for ESA 1, 2, and 3, respectively. Therefore, the observed ratios are higher and show stronger energy progression than the laboratory values.

The simulated flux from the WTPM is obtained with a specific, time-dependent ionization model developed in a series of papers (Bzowski et al. 2013; Sokół & Bzowski 2014; Sokół et al. 2019a, 2020). The main factor ionizing ISN helium atoms is photoionization. To evaluate the importance of ionization on normalization factors, we use the WTPM to calculate expected ISN helium fluxes with the ionization processes switched off. With these simulations, we calculate the ratio of the summed simulated signal over the angular bin included in the study with and without the ionization losses. This ratio is calculated separately for each observational season and it represents the mean survival probability of ISN helium atoms. The black line in the top panel of Figure 3 shows this ratio. Within the current model, the mean survival probability was ~65% during the minima in 2009 and 2020, and it decreased to ~44% in 2015, in the proximity of the solar maximum.

A possible explanation of the observed variability in the RV-norm coefficients is that the actual ISN helium ionization rate differs from that adopted in the WTPM. We consider three possible scenarios with different modifications to the ionization rates that would explain the discrepancy. In the first scenario, we assume that the mean values of the scaling function coefficients over periods with the nominal and the lowered PAC voltage represent the actual instrument efficiency of the instrument. Thus, the actual mean survival probability is given by a product of the WTPM survival probability and the ratios of the best-fit coefficients for each year to these mean values. This scenario is shown in Figure 3 using red bands.

Assuming that the ionization rate at 1 au $\beta$ is constant over time and decreases inversely proportionally to squared distance from the Sun, the survival probability is given as:

$$s = \exp\left(-\beta \int \frac{1}{r(t)^2} dt\right), \tag{15}$$

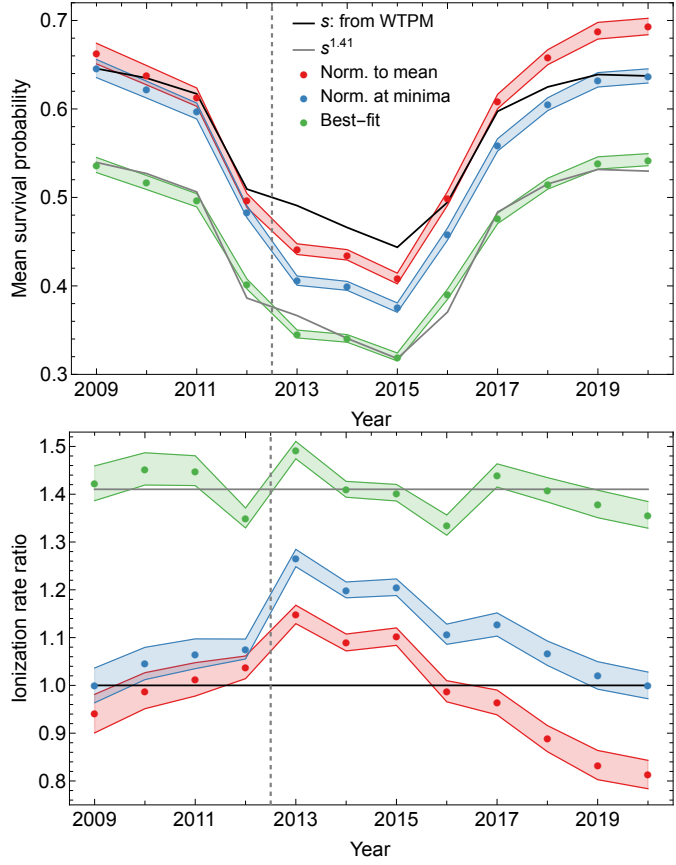

**Figure 3** – Top panel: Mean survival probability obtained from the WTPM with the nominal ionization model (black line) and a proportionally scaled ionization (gray line, see text). Colored lines show mean survival probabilities based on the coefficients of the scaling function obtained under three possible scenarios (see text). Bottom panel: ratio of the ionization rates expected under these scenarios to the ionization rate used in the model. The ratio equal 1 indicates consistency with the adopted ionization model. The vertical dashed line indicates the change in the PAC voltage.

where the integral is over the transport time from the outside of the heliosphere to 1 au. With the survival probabilities $s_1$ and $s_2$ obtained with the ionization rate $\beta_1$ and $\beta_2$, respectively, the ratio of the ionization rates for a given trajectory is, therefore, given by the ratio of the logarithms of the survival probabilities:

$$\frac{\beta_2}{\beta_1} = \frac{\log s_1}{\log s_2}. \tag{16}$$

The same relation holds, assuming that the ratio of the ionization factors is constant even if the actual rates vary over time. The ionization rate ratio that would explain the discrepancy in this scenario is shown with the red line in the bottom panel of Figure 3. This ratio varies over time between ~0.8 and ~1.15. Therefore, this scenario suggests that the ionization rate differ by up to ~20% from the adopted ionization model.

The above scenario predicts that the survival probabilities in the two minima in 2009 and 2020 are significantly different, which is not likely since the ionization proxies are similar in these two minima (Sokół et al. 2019a). Therefore, in the next scenario, we assume that the scaling function coefficients found for minima in 2009 and 2020 are the actual efficiencies of the instrument for the nominal and lower PAC voltage, respectively. The expected mean survival probabilities and ionization rate ratios are shown in Figure 3 using blue lines. Under this scenario, the ionization rate during the maximum should be up to ~25% higher than the ionization rate in the WTPM. Such an increase in the ionization rate would bring the amplitude closer to the amplitude seen in the previous solar cycle (Sokół et al. 2019a). Originally, the ionization rate in the most recent solar cycle is the lowest of the three solar cycles covered in the model.

The last scenario that we consider here assumes that the ionization rate is uniformly increased by a multiplicative factor $\alpha$ over the entire period. Equation (15) indicates that the survival probability with such an increased rate is $s^\alpha$, where $s$ is the survival probability without the increase. In this scenario, we use two unknown coefficients representing the actual instrument efficiency for the nominal and lower PAC voltages: $\epsilon_N$ and $\epsilon_L$. Finally, we find the values of these parameters by minimizing the following sum:

$$\chi^2 = \sum_{y=2009}^{2020} \frac{\left(s_y^{\alpha-1}\epsilon_{\text{PAC}(y)} - a_y\right)^2}{\sigma_{a,y}^2}, \quad (17)$$

where $\text{PAC}(y)$ gives N and L for years with nominal and lowered PAC voltages, respectively. The WTPM modeled mean survival probabilities are given by $s_y$. The minimization of this formula gives best-fit parameters: $\alpha = 1.41 \pm 0.05$, $\epsilon_N = (21.9 \pm 0.6) \times 10^{-6}$ cm$^2$ sr, and $\epsilon_L = (11.2 \pm 0.4) \times 10^{-6}$ cm$^2$ sr. The mean survival probability $s^{1.41}$ obtained with a 41% higher ionization rate using the existing WTPM result $s$ is shown in Figure 3 using the gray line. The survival probabilities estimated from the scaling function coefficients are given as $sa_y/\epsilon_{\text{NL}(y)}$ and are shown with the green line. This significantly larger ionization rate explains most of the discrepancy between the individual scaling function coefficients. The remaining small variation (~5%) is shown with the green line in the bottom panel of Figure 3.

All three scenarios discussed above suggest that a significant modification to the ionization model is required to explain the variations in the scaling function over the entire time period. Since photoionization accounts for at least 80% of the total ionization rate, systematic effects in the calculation of this rate are the first suspect here. In the approach used in the ionization model, the photoionization rates are calculated at Carrington rotation period time resolution based on daily photoionization rates computed directly from the solar spectra measured by TIMED-SEE (Woods et al. 2005). These rates are computed for the years 2002–2016, i.e., for the time interval when the absolute calibration is considered by the authors of the measurements as not affected by aging effects (Woods et al. 2018). For the epochs before 2002 and after 2016, a proxy model is employed, based on Carrington rotation period averages of the F10.7 solar radio flux observations (Tapping 2013) and a nonlinear correlation formula developed by Sokół et al. (2020) using a methodology developed by Sokół & Bzowski (2014). The basis for development of this latter formula was the reaction cross section by Verner et al. (1996) and daily photoionization rates calculated based on observations of the solar EUV spectrum by the TIMED mission (Woods et al. 2005, Data level 3, release 12), with instrument aging effects compensated for (Woods et al. 2018). Usable TIMED data cover an interval from ~2002 until ~2016. While they do not cover the time intervals when the solar activity was the highest, like, e.g., during the solar maxima of 1958 and 1985, ISN helium atoms are most effectively ionized within less than a year before detection. Therefore, even for the earliest IBEX-Lo observations, the solar activity levels were safely within the limits of the solar activity covered by direct TIMED observations.

Below, we discuss possible sources of a systematic bias in the estimates of the ionization rate. A systematic bias resulting from fitting the F10.7 proxy seems unlikely. Calibration of the F10.7 flux is stable within ~2% (Svalgaard 2016), and fits relating the photoionization rate to the F10.7 flux do not show large systematic residuals. Hence, the suspect may be the ionization rate obtained from integration of the TIMED spectra. Topics to address here are the absolute calibration, the uncertainties of the observed spectral flux, and the uncertainty of the cross section for photoionization.

Data available in Version 12 of the TIMED data have two wavelength-dependent relative uncertainties: due to measurement errors, fluctuating between ~1% and 5%, and a "total uncertainty" of ~10%, described in data release notes as "useful for comparing against other measurements and models"[†]. Thus, the total systematic uncertainty of the photoionization rate is expected to be at most 15%.

A recent correction of the spectra for instrument aging (Woods et al. 2018) resulted in a small systematic revision of the previously used helium photoionization rates, as presented in Figure 8 in Sokół et al. (2020). The magnitude of this revision is solar activity-dependent, and it is largest for high activity levels in 2015, by almost –10%. For low solar activity times, as in 2009 and 2019, no correction was necessary. Moreover, the correction is downward, not upward, as it would be needed based on the results of our analysis. Without the correction, the ionization rate of He would feature a secular change in its ratio to solar-activity proxies such as the F10.7 and F30 radio fluxes or the Mg II c/w ratio, i.e., this would suggest a physical change at the Sun affecting the relation between the EUV and radio flux. To our knowledge, such a change has not been reported, which makes us believe that the correction is justified.

Another potential error source is the cross section for photoionization. Sokół et al. (2020) used that suggested by Verner et al. (1996). These authors claim that the accuracy of the original data for the cross section for He I in the low-energy range by Samson et al. (1994) is better than 2%, and 10% for high energies, and that the fitting did not introduce any larger error. Because of the wavelength dependence of the photoionization cross section on one hand, and the shape of solar spectrum in the EUV domain on the other hand, the waveband mostly responsible for ionization of He is ~15–40 nm, which is at the low-energy end of the ionizing spectrum. Thus, uncertainty in the photoionization rate of He because of the uncertainty of the cross section is expected to be only a few percent. In all, a systematic error in the photoionization rate is expected to be at most ~20%, in the unlikely case when all systematic uncertainties sum up algebraically. On the other hand, this magnitude is similar to the increase of ~21% in the solar radiation pressure, found by Rahmanifard et al. (2019) from IBEX-Lo ISN hydrogen observations, relative to the expected pressure from models basing on the Ly-α observations at wavelength ~122 nm.

The other two ionization reactions contribute only ~20% of the total ionization rates, so they would need to account for 100% uncertainty of their rates to be able to add the missing 20%. Therefore, we do not believe that the currently known balance of the uncertainties of the ionization rate of helium sums up to a systematic error of 40% needed to explain the discrepancy under the scenario of the proportional scaling of the helium ionization rate. On the other hand, the above discussed uncertainties of the ionization rate are mostly from systematic effects and thus are not likely to yield error of ~20% required for the scenario with increased ionization only in the maximum.

Potentially, an additional unaccounted signal in the IBEX observations might also explain this discrepancy. If there is some additional flux observed during the minima but suppressed in the maximum, this may result in the observed pattern of the scaling function coefficients. For example, the contribution of ISN hydrogen is only expected in the proximity of solar minima (Galli et al. 2019; Rahmanifard et al. 2019). However, the observed fluxes of ISN hydrogen are too small to introduce a ~20% change in the combined flux. Therefore, this possibility is not likely. Moreover, such contribution would likely skew the resulting ISN helium parameters obtained from fits to individual observational seasons. Such a skew is not observed (see Section 7).

The observed discrepancy in the scaling function representing the instrument efficiency is not fully understood at present. This section briefly discussed the possibility of actual changes in the instrument response, insufficient ionization rates of ISN helium in the model, and the presence of additional signals. However, the magnitude of this effect is large, and each of these explanations separately would require changes larger than the expected uncertainties of the respective effects. Nevertheless, combination of all effects may explain the discrepancy. In Section 7 we show that the inflow parameters derived from individual observational seasons, differently than the scaling parameters, are not correlated with observation time, thus it is not likely that this scaling coefficient discrepancy significantly affects the ISN helium parameters sought in this study.

---

[†] TIMED SEE Version 12 Data Product Release Notes, October 31, 2017, by D.L. Woodraska and T.M. Woods http://lasp.colorado.edu/data/timed_see/SEE_v12_releasenotes.txt

**Table 7**
**Results of the fitting to individual seasons with RV-norm**

| Season | $N$ | $K$ | $\nu$ | $\chi^2$ | $\chi^2/\nu$ | $\lambda$ (°) | $\beta$ (°) | $v$ (km s$^{-1}$) | $T$ ($10^3$ K) |
|---|---|---|---|---|---|---|---|---|---|
| 2009 | 90  | 10 | 80  | 168.8 | 2.110 | 256.3±1.6 | 5.01±0.26 | 25.6±1.5 | 7.3±0.8 |
| 2010 | 72  | 10 | 62  | 99.7  | 1.607 | 252.9±3.6 | 5.19±0.24 | 28.5±3.8 | 9.3±2.6 |
| 2011 | 24  | 6  | 18  | 9.7   | 0.539 | 259.4±1.1 | 4.50±0.21 | 23.9±0.9 | 5.4±0.6 |
| 2012 | 60  | 10 | 50  | 56.2  | 1.125 | 256.7±0.9 | 5.16±0.18 | 25.5±0.8 | 7.3±0.5 |
| 2013 | 180 | 10 | 170 | 250.3 | 1.472 | 254.9±1.0 | 5.05±0.20 | 26.0±0.8 | 7.7±0.5 |
| 2014 | 144 | 10 | 134 | 136.8 | 1.021 | 257.1±0.9 | 4.93±0.21 | 25.1±0.7 | 6.7±0.4 |
| 2015 | 96  | 10 | 86  | 87.5  | 1.017 | 253.0±2.0 | 5.12±0.30 | 28.7±1.8 | 9.2±1.5 |
| 2016 | 36  | 8  | 28  | 29.2  | 1.043 | 259.8±4.4 | 4.98±0.57 | 23.1±3.1 | 6.1±2.3 |
| 2017 | 120 | 10 | 110 | 157.8 | 1.434 | 255.2±0.7 | 4.88±0.22 | 26.9±0.7 | 8.1±0.4 |
| 2018 | 144 | 10 | 134 | 206.7 | 1.542 | 255.5±0.5 | 5.00±0.27 | 26.8±0.4 | 7.9±0.3 |
| 2019 | 144 | 10 | 134 | 208.2 | 1.554 | 256.8±0.7 | 5.12±0.20 | 24.9±0.6 | 7.0±0.4 |
| 2020 | 144 | 10 | 134 | 210.6 | 1.571 | 254.1±0.8 | 5.13±0.19 | 26.6±0.7 | 8.2±0.5 |

## 7. Temporal Stability of ISN Helium Parameters

In this study, we analyze IBEX-Lo ISN data from 12 observational seasons between 2009 and 2020. This period covers the entire solar cycle 24, and thus we can examine whether our results are affected by the solar cycle. This 12-year long period provides insight into hypothetic decade-long systematic changes of the interstellar flow determined based on 1 au observations.

Each observational season from IBEX may be independently analyzed using the formalism presented in Section 4. When only one season is considered, there is no difference between using scaling functions in the form of YV-norm, RV-norm, or PV-norm (see Table 4). The best-fit parameters from individual observational seasons are provided in Table 7. The scatter in the best-fit parameters is clearly visible. The 1σ uncertainties provided in the table do not provide a full description due to substantial correlations between fit parameters. Correlations between selected pairs of the inflow parameters are shown in Figure 4 as 1-sigma ellipses representing the correlated uncertainties of the parameter pairs. The scattering of the best-fit results for individual years approximately follows the directions indicated by elongation of the ellipses, which correspond to the IBEX correlation tube. Even though the correlation tube is not explicitly included in the analysis, it is recovered by the covariance matrix.

Most of the results obtained for individual observational seasons overlap with the ellipse showing the global best-fit parameters. However, the result from the year 2011 is significantly off from the rest of the results. This is the only year in which we have observational data from only one of the ESA steps collected in the special mode (see Table 1 and 2). As shown by Swaczyna et al. (2018), results from individual ESA steps may not correctly reconstruct the relative instrument response function. Consequently, this result is likely biased if interpreted independently from the remaining data. Note that the reduced $\chi^2/\nu$ in this case is much smaller than in all other years.

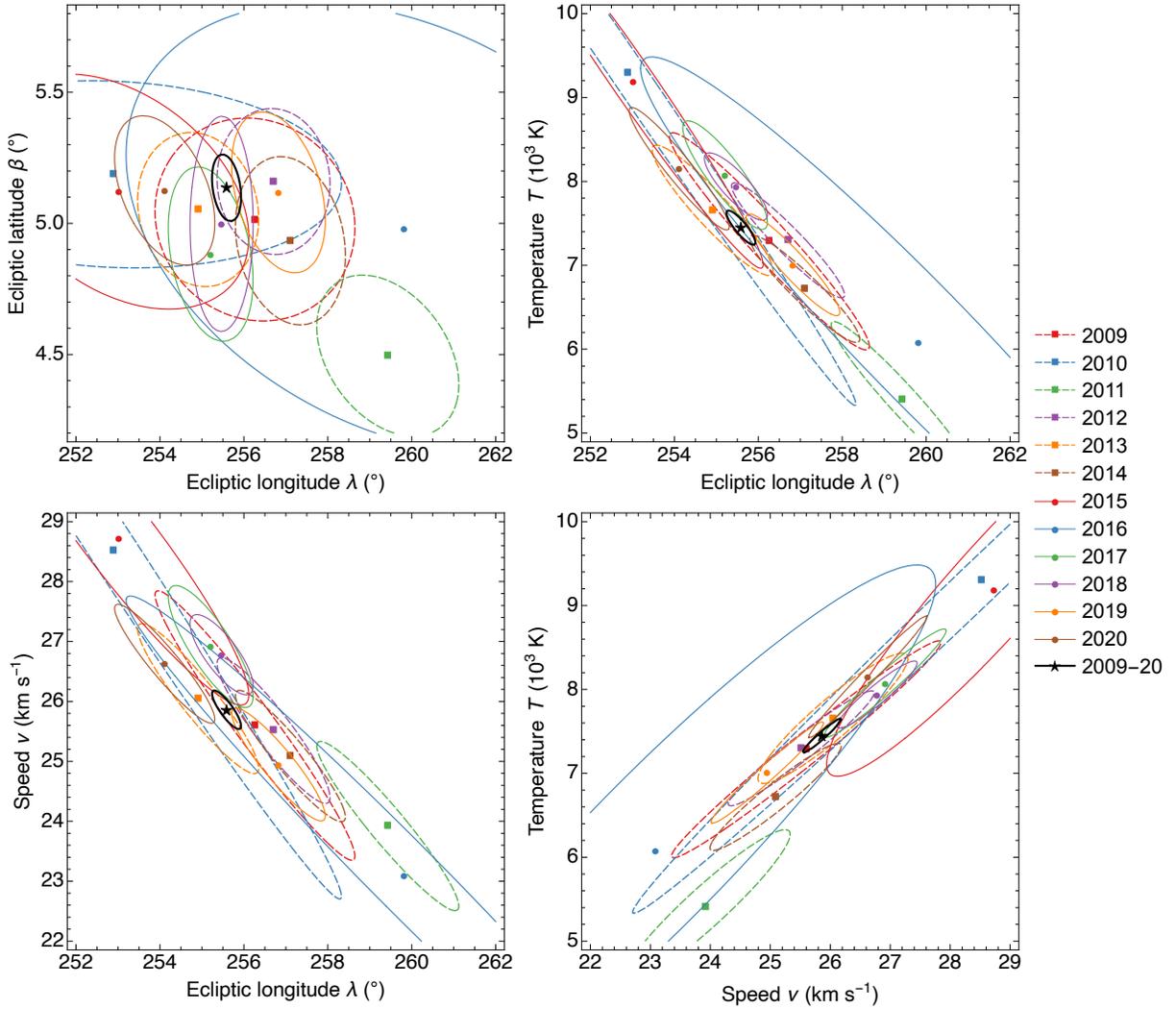

**Figure 4** – Fitting results to individual observational seasons (color symbols) and all seasons (black symbols). The best-fit parameters are marked with filled symbols, and ellipses indicate 1σ uncertainty ranges. The following pairs of parameters are presented in panels: longitude-latitude (top left), longitude-temperature (top right), longitude-speed (bottom left), and speed-temperature (bottom right).

We performed statistical tests of the hypothesis that there exists an unspecified temporal change in the inflow velocity and the temperature as listed in Table 7. To that end, we tested the hypothesis that the parameters are correlated with time. Different tests are needed for different types of correlation, which need not be linear or even monotonic. For the time series of parameters listed in Table 7, we performed tests available in the Mathematica 12 software package. They cover different types of correlation and include the Blomqvist β, Goodman-Kruskal γ, Hoeffding D, Kendall τ, Pearson correlation, Pillai Trace, Spearman Rank, and Wilks W statistical tests. All of these tests suggest at the significance level 0.01 that we must not reject the hypothesis that the parameters are uncorrelated with time. Therefore, even if a change of parameters occurred in the VLISM, it is too small to be reflected in parameters obtained from individual observation seasons.

Since time-changes in the parameters of the ISN helium flow, if they exist, must be small, we perform an additional test in which each of the parameters is allowed to slowly change over time. The first-order approximation of such change is that each parameter changes linearly as a function of observation season year ($y$):

$$\alpha = \alpha_0 + x_\alpha(y - 2014.5) \qquad (18)$$

where $\alpha = \lambda, \beta, v, T$ represents each of the ISN helium parameters. We use the mean observation season covered in the study $y = 2014.5$ as a reference point, to minimize correlation between the mean value $\alpha_0$ and the linear coefficient $x_\alpha$. We include the formula describing the change of parameters (Equation 18) in the interpolation of the simulations given in Equation (6). The global fit is repeated with 4 more fit parameters than the global fit without time dependency. We use the RV-norm form of the scaling function.

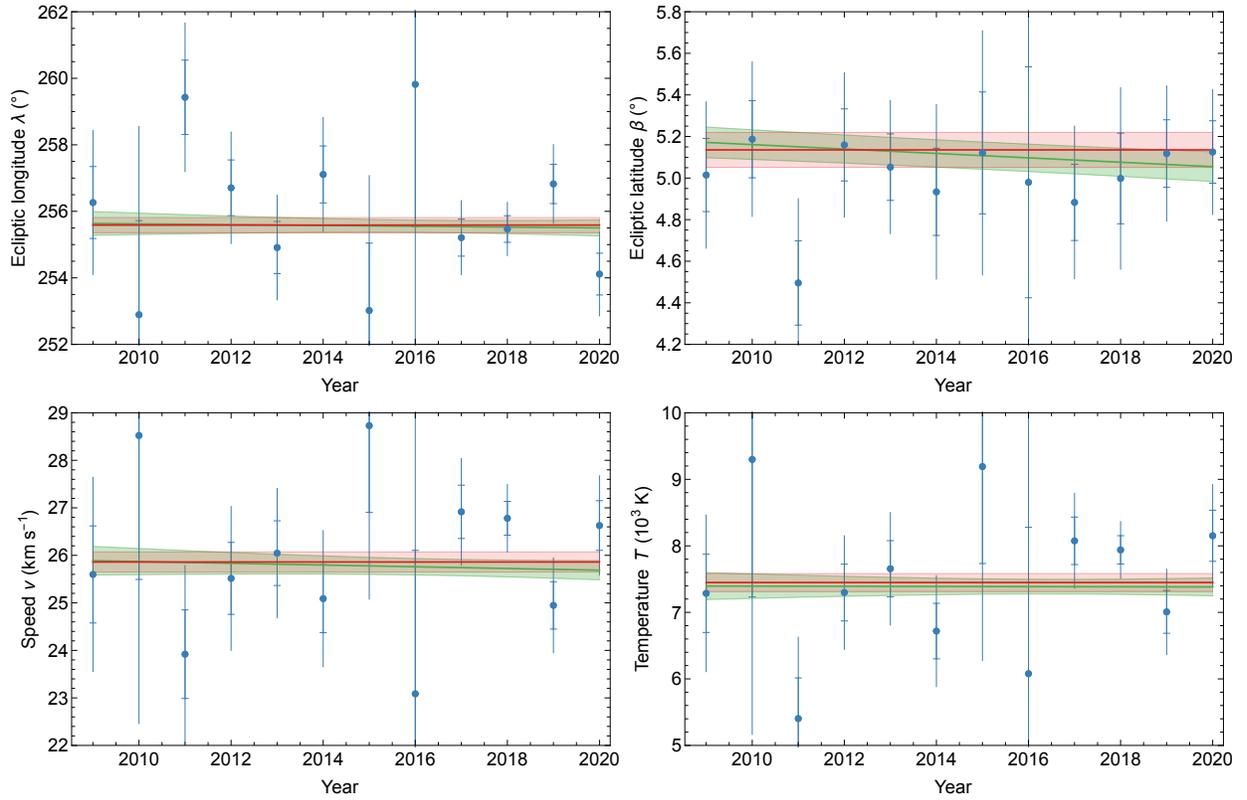

**Figure 5** – Temporal changes in the best-fit parameters of the ISN helium flow: longitude (top left), latitude (top right), speed (bottom left), and temperature (bottom right). Blue symbols show the results of independent fits to each observational season. The bars show 2σ uncertainties, and 1σ uncertainties are limited by fences. The red line presents the global time-independent fit (see Section 5) with the 1σ uncertainty band, and the green line shows the fit with linear changes in each parameter also with 1σ uncertainty band. Note that these bands overlap in large portions of the plots.

The best-fit values of the time-dependent coefficients are $x_\lambda = (-0.012° \pm 0.051°)/\text{yr}$, $x_\beta = (-0.011° \pm 0.007°)/\text{yr}$, $x_T = (-0.001 \pm 0.027) \times 10^3$ K/yr, $x_v = (-0.018 \pm 0.039)$ km s$^{-1}$/yr. Figure 5 shows the best-fit dependency of the fit parameters with green line. Only the change in the ecliptic latitude exceeds its 1σ uncertainty. It is likely that this is purely statistical fluctuation, as we normally expect that the probability that the true value is outside the 1σ uncertainty range is ~32%. Moreover, this apparent time-dependency may result from the spin phase shift as discussed in Appendix B. Therefore, this coefficient likely does not represent an actual change in the ISN helium flow, rather it is a statistical fluctuation or possibly an instrumental effect. The changes in all other parameters are within 1σ uncertainty ranges, and the best-fit coefficient of the change in the longitude is by order of magnitude smaller than (0.14° ± 0.06°)/yr reported by Frisch et al. (2015). Therefore, we conclude that the IBEX observations do not indicate any decade-long changes in the flow direction.

The time changes included in Equation (18) corresponds to the IBEX observation season, and therefore describe the time-evolution of the ISN helium parameters derived at a given time from observations at 1 au, similarly to what was done previously by Frisch et al. (2013, 2015). To reconstruct possible changes in the pristine VLISM, one would need to account for the transport time and its variation from the VLISM to 1 au (Bzowski & Kubiak 2020). However, gradient limits obtained in this study indicate that such changes must be small and a longer observation period is needed to address physical changes in the VLISM.

## 8. Conclusions and Summary

IBEX-Lo has measured ISN helium atoms over a full solar cycle. In this study, we analyze 12 observational seasons of the ISN helium from 2009 to 2020, extending between two minima of solar activity. The data statistics have been significantly improved compared to the previous set of analyses of the IBEX-Lo ISN data (Bzowski et al. 2015; McComas et al. 2015b; Schwadron et al. 2015; Swaczyna et al. 2018; Wood et al. 2019), which used the data collected only from 2009 to 2014. The data coverage is further improved thanks to enhanced accumulation times of the relevant energy steps in the most recent years (see Table 1). Consequently, the total number of observed counts during the ISN seasons increased three-fold from ~0.6 million cumulatively between 2009 and 2014 to ~1.8 million cumulatively between 2009 and 2020 (see Table 2). We analyze observed data by comparing them with the simulations obtained from the WTPM (Sokół et al. 2015b).

The best-fit parameters of the ISN helium at the heliopause obtained in this study are the inflow direction $(\lambda, \beta) = (255.59° \pm 0.23°, 5.14° \pm 0.08°)$, speed $v_{HP} = 25.86 \pm 0.21$ km s$^{-1}$, and temperature $T_{HP} = 7450 \pm 140$ K. These results are almost identical (see Table 6) with the one reported by Bzowski et al. (2015), but the parameter uncertainties are decreased about two-fold, except for the inflow ecliptic latitude in which systematic factors dominate the final uncertainty. Even though the reported uncertainties have been scaled due to a too high best-fit $\chi^2$ (see Section 4), they do not account for bias from unidentified systematic sources of the IBEX signal. Accounting for change between the pristine VLISM and the heliopause due to attraction in the solar gravitational field (McComas et al. 2015b) and elastic collisions in the outer heliosheath (Swaczyna et al. 2021), the pristine VLISM speed and temperature are $v_{VLISM} = 25.9$ km s$^{-1}$, and $T_{VLISM} = 6150$ K.

The coefficients of the scaling functions used to model the instrument energy response show a significant time-evolution over the solar cycle. While during the mission, the instrument efficiency was reduced approximately two-fold when the PAC voltage was decreased in 2013, it should remain constant before and after this change. However, the obtained coefficients show significant increase during the period with the lower PAC voltage. The most straightforward interpretation that the instrument efficiency changes over time seems unlikely since it suggests decreasing efficiency over the first four years (coinciding with the period of the nominal PAC voltage) and increasing by ~20% efficiency over the later years (the lower PAC voltage). Since the pattern of this time-evolution appears to be correlated with the solar cycle phase, we also considered a possibility that ionization rates used to model the ISN helium losses are underestimated. We find that by increasing the ionization rate up to ~25% during the solar maximum or by increasing the ionization rate proportionally by ~40% over the entire period, we explain most of the observed time-evolution of the scaling function coefficients. Both these deviations are, however, much stronger than the expected uncertainty ranges of the ionization rate estimation. Alternatively, some additional signals not included in this study may contribute to the observed discrepancy. In summary, we are not able to settle which hypothesis is responsible for this evolution. It is likely that more than one of the considered effects contribute to the observed pattern. The problem can be further investigated with other observational data of pickup ions and ionization processes.

In addition to the global fit to the entire IBEX dataset, we also perform fits to individual observational seasons. We verified that the scatter of these individual results is mainly along the IBEX correlation tube, and there is no indication of a systematic pattern in the time evolution of other parameters. We do not see any indication of long-term evolution of the ISN helium parameters estimated at 150 au over the covered period of the IBEX mission. Therefore, the observed changes in the scaling function coefficient do not impact our estimation of the ISN helium parameters.

The IBEX-Lo observations of the ISN helium show that direct sampling of ISN atoms with a low background instrument is crucial to determine the VLISM conditions ahead of the heliosphere. IBEX-Lo observations are limited by the fixed boresight direction relative to the IBEX spin axis, allowing only detection of atoms near their perihelia and resulting in the IBEX correlation tube. This limitation will be removed on Interstellar Mapping and Acceleration Probe (IMAP; McComas et al. 2018) with a pivot platform allowing for adjusting the elevation angle between the spacecraft spin axis and the IMAP-Lo boresight. This change gives a flexible observational strategy allowing for sampling of ISN atom populations over the larger part of each year (Sokół et al. 2019c), including observations of indirect trajectories of the ISN helium flow. Such observations provide direct insight into the total ionization rates for He, and this will shed light on the hypotheses of the modified ionization presented in this study. Effectively, this strategy results in crossing of correlation tubes narrowing the uncertainty region of the ISN helium parameters (Schwadron et al. 2022).

The IBEX-Lo operation should be continued to enable cross-calibration with IMAP-Lo and give continuous observation over at least 20 years. Even a longer period, exceeding three solar cycles, is desired to observe possible physical changes in the pristine VLISM (Bzowski & Kubiak 2020). The VLISM is the only interstellar medium that can be directly sampled, and therefore direct sampling experiments give a unique insight into properties of the interstellar medium that are not measurable with telescopic observations of absorption spectra of interstellar material between the Sun and the nearest stars.

*Acknowledgments*: This work was funded by the IBEX mission as a part of the National Aeronautics and Space Administration (NASA) Explorer Program (80NSSC20K0719), and by IMAP as part of the Solar Terrestrial Probes Program (80GSFC19C0027). P.S. and F.R. acknowledge support by the NASA under Grant No. 80NSSC20K0781 issued through the Outer Heliosphere Guest Investigators Program. M.B. and M.A.K. were supported by Polish National Science Centre grant 2019/35/B/ST9/01241.

## Appendix A. Quiet TOF 2 Times

The TOF 2 rates are collected over six 60° wide sectors and are accumulated over 64 spins, during which each ESA step is observed for 8 spins if the standard ESA sweep table (1-2-3-4-5-6-7-8) is used. To have the same "good times" for all ESA steps used in this study, we only use TOF 2 rates for ESA step 2 in this criterion. In the case of the doubled ESA coverage (ESA sweep tables 1-1-2-2-5-6-7-8 and 1-1-2-2-3-6-7-8), the TOF 2 monitor rates are stored in two histograms, each covering 8 spins, and for each sector bin, we select the mean value from these two histograms. During the Oxygen and X-cal modes, ESA step 2 is accumulated over the entire 64 spins, and the special mode ESA setting is subsequently accumulated over 7×64 spins. In this case, the ESA step 2 TOF 2 monitor rates are stored to 8 histograms, and we also take the mean of these histograms in each sector further in the procedure.

As mentioned above in Section 2.2, the sector with the ISN flow have much higher rates and are not used. Therefore, we take the two neighboring sectors, and we choose the larger of these two. This selection is made after taking the mean from multiple histograms for the non-standard ESA sweep table as described in the previous paragraph. This procedure gives a single time series with a time step corresponding to 64 spins (non-special modes) or 512 spins (special modes). Only the periods for which the obtained rate is below some threshold value (see the next paragraph) are considered to meet the TOF 2 criterion. Additionally, for the rates with the 64-spin time resolution, we require that the moving mean over 3, 5, and 7 consecutive time bins is also below the threshold. This criterion cut out some bins on the edges of the good periods. However, if a single time bin that does not meet this requirement is surrounded by bins that do meet the criterion, this time bin is included in the good time list. The moving mean criterion is not applied to the data with 512-spin time resolution.

The threshold value for the time series is obtained as a larger value from the following two:

1. 1.6 times the 15$^{th}$ percentile of the rates in the time series for a considered orbit after removing time bins with values below 10 and above 300.
2. 2 times the median value of the rates from all orbits after removing the time bins with the values below 10 and above 300. This threshold is, in practice, equal to 84.

The above criteria are tuned so that the "good time" selections approximately match the previous estimation of the quiet TOF 2 times from previous studies of the IBEX-Lo data (Leonard et al. 2015; Möbius et al. 2015a). Formerly, this criterion was manually applied to the data from each orbit, while the presented algorithm allows for full automation of the process. Moreover, this threshold provides a reasonable balance of the data that meet this criterion but do not introduce significant background. Rates below 10, which are excluded from finding the threshold, indicate that the data are likely not complete, and we remove such times from our study. Rates above 300 indicate extremely high background and are also removed. The mean threshold is 70 counts, and for all orbits, it is between 37 and 84 counts.

## Appendix B. Spin Phase Offset Since Orbit 326

Initial analysis of the ISN observations discussed here revealed that the ecliptic latitude of the ISN helium flow is systematically shifted by ~0.6° for observational seasons 2017-2020. The best-fit latitude to individual seasons between 2009 and 2016 was between 4.72° and 5.06°, while for the seasons between 2017 and 2020, the latitude was found between 5.47° and 5.87°. Moreover, the fits with these two subsets gave $\chi_\nu^2$ of 1.88 and 2.69, respectively, but the fit to the entire dataset gave $\chi_\nu^2 = 6.52$. Such a drastic increase in the reduced $\chi^2$ suggests that the two subsets are not compatible.

The analytic analysis method of the IBEX-Lo observations (Lee et al. 2012, 2015) shows a direct connection between the inflow ecliptic latitude and the position of the signal peak for a given orbit in the spin angle of the spacecraft. The histograms used in this study are organized based on the spin phase, which should be triggered by the spacecraft ACS 3° before the IBEX-Hi boresight points to the north ecliptic pole (NEP). In 2016, the spacecraft ACS was reprogrammed due to the star tracker anomaly caused by the degradation of the system. The star tracker anomaly occurred early in orbit 315b. The star tracker was off until orbit 326 (except for 2 brief test intervals in 325b). Based on this presumptive clue, we suspect that the spin phase may be incorrectly triggered with some offset after the software change.

The IBEX-Lo team anticipated that the knowledge of the spin phase is crucial for studies of the ISN data. However, the desired precision exceeded the anticipated performance of the ACS. For this reason, a star sensor was aligned with the IBEX-Lo boresight to support the determination of the spin axis and spin phase. The spacecraft orientation derived from the star sensor observations was verified to be in very good agreement with the ACS data based on the data obtained during the first two years of the mission (Hłond et al. 2012). However, due to a limited number of bright stars in the sensor field of view, the star sensor data are not routinely used to determine the spacecraft orientation. Nevertheless, since the star sensor gives independent information on the spacecraft orientation, we use these data to verify if the changes made in 2016 affected the triggering of the spin pulse.

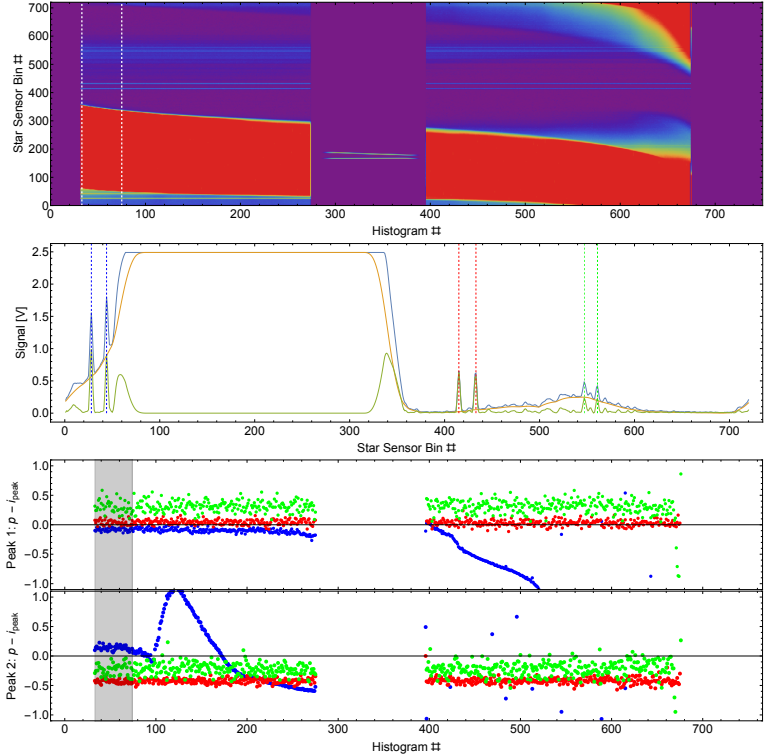

**Figure B1** – Star sensor data reduction. The top panel shows a 2-dimensional histogram of measured star sensor voltages with the histogram number on the horizontal axis and the bin number on the vertical axis. The middle panel presents the averaged star sensor data over the period between the two dashed lines in the top panel. The blue, orange, and green lines show the raw signal, the estimated background, and the signal with background subtracted. The vertical color lines show the local maxima of the signal. The bottom panels show the peak positions relative to the peak bin for the three stars identified in this data and the time period used for this orbit (gray shaded area).

## B.1 Star Positions from Star Sensor Data

The star sensor aperture consists of two slits inclined at an angle to each other to enable the determination of the elevation relative to the spacecraft spin plane (Hłond et al. 2012). Consequently, stars appear as two peaks in star sensor histograms. The top panel of Figure B1 presents a 2-dimensional representation of star sensor data from orbit 120. The horizontal axis shows the histogram number organized according to the time, but note that histograms are typically averaged over multiple spins. The vertical axis corresponds to spin bins in which the star sensor data are stored. Unfortunately, the star sensor measurements are often saturated by the Moon or Earth's glow (appears as large red areas in this plot).

For each analyzed orbit/arc, we select a period maximizing the number of observed stars. In the example used in this figure, we select the short part soon after the beginning of the orbit, when three stars can be identified. Stars are observed against a comparably strong diffuse background, which may influence the determination of the positions. Therefore, we estimate the background using the EstimatedBackground function in Mathematica with a scale factor of 6 for each histogram separately. This function smooths all features extending over more than 6 bins. This background is subsequently subtracted from the star sensor data. This allows for the removal of slowly evolving backgrounds.

The star sensor signal averaged over the selected period is presented in the middle panel of Figure B1. This average is used to find the local maxima in the data. The vertical dashed lines show pairs of local maxima corresponding to three stars identified in this orbit. Since the stars extend over more than one bin, we can determine the centroid (peak position) of each peak using the following formula:

$$p = \frac{\sum_{i=i_{\text{peak}}-3}^{i_{\text{peak}}+3} i s_i}{\sum_{i=i_{\text{peak}}-3}^{i_{\text{peak}}+3} s_i} \tag{B1}$$

where $i$ enumerates the bin numbers, $i_{\text{peak}}$ is the bin number with the local maximum, and $s_i$ is the measured star sensor voltage with the background subtracted in bin $i$. Note that while the peak bin $i_{\text{peak}}$ is an integer bin, the peak position allows us to better estimate the star position using neighboring bins. The differences between the peak position and the peak bin indicate where each star is located relative to this bin and should normally be within ±0.5. The peak positions relative to the peak bin are presented in the two bottom panels of Figure B1. Finally, the peak positions are averaged over the selected period (a gray shaded area in these panels).

Voltage measured by the star sensor is sampled in 720 bins covering the full spin. The star sensor bin width is predetermined based on the spin period and corresponds to an integer multiplication of a base time tick of $t_i = 1/14{,}400$ s (Hłond et al. 2012). This integer is obtained as $k = \left\lceil \frac{T_{\text{spin}}}{720 t_i} \right\rceil$, where $T_{\text{spin}}$ is the spin period, and $[z]$ is the ceiling integer of a real number $z$. Consequently, the bin width in degrees is

$$\gamma = \frac{k t_i}{T_{\text{spin}}} 360°. \tag{B2}$$

This scheme provides that the entire spin is stored in no more than 720 bins. If fewer bins cover the entire spin, the remaining bins are not updated, and the values reported in these bins are incorrect. Nevertheless, we do not identify any stars in the proximity of these bins, and thus this issue does not affect our result.

The voltage is sampled at the beginning of each bin, but due to a delay introduced by an amplifier, with a time constant of 12.5 ms, the star position is shifted by 0.314° for a typical spin period of 14.34 s. Even though the spin period slightly changes over time, the variation of this shift is negligible. Therefore, a peak in the star sensor bin $p$ corresponds to the North Ecliptic Pole (NEP) angle

$$\alpha = \gamma(p - 1) + 180° - 3° - 0.314°, \tag{B3}$$

where we use a convention in which bins are numbered from $p = 1$ to $p = 720$. Since the spin pulse is emitted 3° before IBEX-Hi points towards NEP, we shift the angle by $180° - 3°$.

For each star, we find two peak positions $p_1$ and $p_2$, which are transformed to NEP angles following Equation (B3): $\alpha_1$ and $\alpha_2$, respectively. These two peak positions give the star position in the spacecraft frame (Hłond et al. 2012 Equations 2-3):

$$\alpha = \frac{\alpha_1 + \alpha_2}{2}, \tag{B4}$$

$$\delta = \arcsin\left[\frac{\tan\frac{\alpha_2 - \alpha_1 - \sigma_V}{2}}{\tan\beta_V}\right], \tag{B5}$$

where $\alpha$ and $\delta$ are the NEP angle and elevation of the star, respectively, $\sigma_V = 8.4°$ is the slit separation at 0° elevation, and $\beta_V = 14.4°$ is the tilt angle of each slit. Later, the position of each star in the spacecraft frame is transformed to equatorial coordinates using the spacecraft spin axis. This position is compared with the 300 brightest stars in the sky, and the closest one within 1° from the star sensor position is selected. If a star is not identified, it is most likely a bright planet, but we left them out from this study. Finally, we transform the catalog position to the spacecraft frame, and compare NEP and elevations angles.

**Table B1**
**Position of Stars Identified in Star Sensor Data for Select Orbit**

| Orbit/Arc | Star Sensor | | | | Star Catalog | | | | Difference | | Star |
|---|---|---|---|---|---|---|---|---|---|---|---|
| | NEP | Elev. | R.A. | Decl. | NEP | Elev. | R.A. | Decl. | ΔNEP | ΔElev | |
| 24 | 194.223 | -0.888 | 96.152 | -52.666 | 194.177 | -0.795 | 95.988 | -52.696 | 0.046 | -0.093 | Canopus |
| 24 | 28.346 | 0.968 | 279.177 | 38.700 | 28.269 | 1.023 | 279.234 | 38.783 | 0.077 | -0.055 | Vega |
| 24 | 93.570 | -2.666 | 283.186 | -26.498 | 93.427 | -2.084 | 283.816 | -26.297 | 0.143 | -0.582 | Nunki |
| 72 | 194.219 | -0.576 | 95.968 | -52.650 | 194.177 | -0.596 | 95.988 | -52.696 | 0.043 | 0.020 | Canopus |
| 72 | 28.330 | 0.668 | 279.230 | 38.722 | 28.270 | 0.679 | 279.234 | 38.783 | 0.060 | -0.011 | Vega |
| 72 | 93.550 | -3.161 | 283.352 | -26.467 | 93.423 | -2.730 | 283.816 | -26.297 | 0.127 | -0.431 | Nunki |
| 120 | 194.211 | -0.611 | 96.017 | -52.664 | 194.177 | -0.599 | 95.988 | -52.696 | 0.034 | -0.012 | Canopus |
| 120 | 28.338 | 0.708 | 279.337 | 38.726 | 28.270 | 0.636 | 279.234 | 38.783 | 0.068 | 0.072 | Vega |
| 120 | 93.531 | -3.252 | 283.454 | -26.443 | 93.419 | -2.914 | 283.816 | -26.297 | 0.112 | -0.338 | Nunki |
| 161b | 194.169 | -0.918 | 95.874 | -52.655 | 194.135 | -0.991 | 95.988 | -52.696 | 0.034 | 0.073 | Canopus |
| 161b | 230.362 | -2.769 | 101.379 | -16.696 | 230.340 | -2.684 | 101.289 | -16.713 | 0.022 | -0.086 | Sirius |
| 161b | 28.253 | 2.360 | 279.466 | 38.723 | 28.182 | 2.183 | 279.234 | 38.783 | 0.071 | 0.177 | Vega |
| 161b | 93.597 | 1.883 | 283.846 | -26.444 | 93.448 | 1.871 | 283.816 | -26.297 | 0.149 | 0.012 | Nunki |
| 201b | 194.146 | -1.260 | 95.875 | -52.664 | 194.119 | -1.331 | 95.988 | -52.696 | 0.026 | 0.071 | Canopus |
| 201b | 230.335 | -3.492 | 101.389 | -16.701 | 230.318 | -3.397 | 101.289 | -16.713 | 0.017 | -0.095 | Sirius |
| 201b | 28.253 | 2.944 | 279.561 | 38.697 | 28.155 | 2.693 | 279.234 | 38.783 | 0.098 | 0.251 | Vega |
| 201b | 93.493 | 2.858 | 284.036 | -26.315 | 93.455 | 2.664 | 283.816 | -26.297 | 0.038 | 0.194 | Nunki |
| 241b | 194.058 | -1.845 | 96.008 | -52.697 | 194.058 | -1.833 | 95.988 | -52.696 | 0.000 | -0.012 | Canopus |
| 241b | 230.170 | -4.986 | 101.140 | -16.746 | 230.206 | -5.127 | 101.289 | -16.713 | -0.037 | 0.141 | Sirius |
| 241b | 28.136 | 3.803 | 279.338 | 38.689 | 28.040 | 3.724 | 279.234 | 38.783 | 0.096 | 0.080 | Vega |
| 281b | 194.052 | -1.862 | 95.958 | -52.697 | 194.054 | -1.880 | 95.988 | -52.696 | -0.002 | 0.018 | Canopus |
| 281b | 230.154 | -5.050 | 101.083 | -16.752 | 230.198 | -5.247 | 101.289 | -16.713 | -0.044 | 0.196 | Sirius |
| 281b | 28.132 | 3.769 | 279.193 | 38.683 | 28.032 | 3.802 | 279.234 | 38.783 | 0.100 | -0.033 | Vega |
| | | | | | | | | Mean | 0.056 | -0.019 | |
| | | | | | | | | Std. Dev. | 0.053 | 0.199 | |
| 362b | 27.717 | 1.109 | 279.410 | 39.339 | 28.255 | 0.914 | 279.234 | 38.783 | -0.538 | 0.195 | Vega |
| 362b | 92.972 | -1.386 | 283.403 | -25.852 | 93.449 | -1.058 | 283.816 | -26.297 | -0.478 | -0.328 | Nunki |
| 402b | 193.599 | -1.197 | 95.716 | -53.248 | 194.168 | -1.284 | 95.988 | -52.696 | -0.569 | 0.087 | Canopus |
| 402b | 229.820 | -1.593 | 101.254 | -17.295 | 230.403 | -1.570 | 101.289 | -16.713 | -0.583 | -0.022 | Sirius |
| 402b | 27.749 | 2.134 | 279.454 | 39.305 | 28.252 | 1.912 | 279.234 | 38.783 | -0.503 | 0.222 | Vega |
| 442b | 193.571 | -1.010 | 95.524 | -53.225 | 194.123 | -1.242 | 95.988 | -52.696 | -0.552 | 0.232 | Canopus |
| 442b | 229.721 | -3.246 | 101.264 | -17.313 | 230.322 | -3.237 | 101.289 | -16.713 | -0.601 | -0.009 | Sirius |
| 442b | 27.641 | 2.837 | 279.544 | 39.314 | 28.160 | 2.570 | 279.234 | 38.783 | -0.519 | 0.267 | Vega |
| 442b | 92.944 | 2.581 | 283.850 | -25.783 | 93.453 | 2.501 | 283.816 | -26.297 | -0.508 | 0.080 | Nunki |
| 482b | 193.518 | -1.277 | 95.707 | -53.269 | 194.102 | -1.405 | 95.988 | -52.696 | -0.584 | 0.128 | Canopus |
| 482b | 229.668 | -3.924 | 101.310 | -17.330 | 230.284 | -3.877 | 101.289 | -16.713 | -0.617 | -0.047 | Sirius |
| 482b | 27.592 | 3.306 | 279.684 | 39.324 | 28.122 | 2.937 | 279.234 | 38.783 | -0.529 | 0.368 | Vega |
| 482b | 92.959 | 2.946 | 283.286 | -25.848 | 93.452 | 3.378 | 283.816 | -26.297 | -0.493 | -0.431 | Nunki |
| | | | | | | | | Mean | -0.544 | 0.057 | |
| | | | | | | | | Std. Dev. | 0.044 | 0.230 | |

### B.2 Star Positions Before and After the Anomaly

Table B1 presents the results obtained for orbit 24 and the orbit with the nearest spin axis pointing in the subsequent years. These orbits are particularly useful since four bright stars are in the scanned region of the sky: Sirius, Canopus, Vega, and Nunki. Typically, we can identify three out of these four stars. Unfortunately, due to background contamination, not all of them are always visible. The table is split into two periods, before and after the star tracker anomaly. The mean difference in the elevation angle between the position determined from the star sensor and the one from the star catalog is consistent with 0 before (–0.019°±0.199°) and after (0.057°±0.230°) the anomaly. However, the difference in the NEP angle shows a systematic effect. Before the anomaly, it is $\Delta NEP = 0.056° \pm 0.053°$, which is almost consistent with no shift. However, after the ACS update, the mean difference is $\Delta NEP = -0.544° \pm 0.044°$, which is significantly different from zero. Therefore, this result confirms that the spin pulse is systematically offset, starting with orbit 326. The best estimation of this shift is the difference of ΔNEP before and after the ACS update: 0.60°±0.07°.

The star sensor data during the core ISN orbits and arcs do not show as many bright stars, or the peaks from these stars significantly overlap. Nevertheless, we verified that these few stars identified on the ISN orbits show a shift in NEP angle consistent with the above-described shift.

Since the spin angle shift changes the boundaries of the spin phase histogram bins collected by the onboard computer, it is impossible to correct the HB data to match the originally planned bin boundaries in further processing of the data. Therefore, the spin angle offset needs to be appropriately included in models used for comparison with the IBEX data.

## Appendix C. Derivative Products

The raw IBEX-Lo data used in this study has been released on the IBEX website: https://ibex.princeton.edu/RawDataReleases. Additionally, concurrently with this paper, we provide derivative products which may be useful for further investigation of IBEX observations. The derivative products have been posted on Zenodo under Creative Commons Attribution license: doi:10.5281/zenodo.5842422.

The following derivative products has been provided in following ASCII files:

1. `NotSpunTimes.dat` – list of periods during the ISN seasons that are not affected by the loss of the spin pulse synchronization (see Section 2.2).
2. `inertial_pointing_ISN.dat` – list of IBEX-Lo spin axis pointing for each orbit/arc.
3. `oXXXXx-Ee.dat` – culled ISN data collected on orbit or arc `XXXXx` in ESA step `e`. These files include the histogram bin (HB) data from IBEX-Lo in all 60 spin bins separated into 512-spin blocks. The begin and end times of each block are provided.
4. `IBEX-Lo-ISN-He.dat` – orbit-accumulated IBEX-Lo ISN helium data. The files include all corrections and background estimation used in this study (Section 2.3), as well as the uncertainty estimations (Section 2.4).